\documentclass[12pt]{article}
\usepackage{amsfonts}
\usepackage{amsmath}
\usepackage{amssymb}
\usepackage{amsthm}
\usepackage{accents}
\usepackage{youngtab}
\usepackage{braket}
\usepackage{graphicx}
\usepackage[svgnames]{xcolor}
\usepackage{ytableau}
\usepackage{here}
\usepackage{comment}
\usepackage{tikz}
 \usetikzlibrary{patterns}
\usepackage{feynmf}
\usepackage{hyperref}
\usepackage{simplewick}

\newcommand{\ba}{\begin{eqnarray}}
\newcommand{\ea}{\end{eqnarray}}

\makeatletter
\newcommand{\douwidehat}[2]{%
  \sbox0{$\m@th#1\widehat{\hphantom{#2}}$}%
  \sbox2{$\m@th#1x$}
  \sbox4{$\m@th#1#2$}
  \dimen0=\ht0
  \advance\dimen0 -.8\ht2
  \dimen2=\dp4
  \rlap{%
    \raisebox{\dimexpr\dimen0-\dimen2}{%
      \scalebox{1}[-1]{\box0}%
    }%
  }%
  {#2}%
}
\makeatother

\makeatletter
\@addtoreset{equation}{section}

\makeatother

\begin{document}
\begin{titlepage}

\begin{flushright}
UT-18-23
\end{flushright}

\vskip 12mm

\begin{center}
{\Large Plane Partition Realization of }\\{\Large (Web of) $\mathcal{W}$-algebra Minimal Models}
\vskip 2cm
{\Large Koichi Harada and Yutaka Matsuo}
\vskip 2cm
{\it Department of Physics, The University of Tokyo}\\
{\it 7-3-1 Hongo, Bunkyo-ku, Tokyo 113-0033, Japan}
\end{center}
\vfill
\begin{abstract}
Recently, Gaiotto and Rap\v{c}\'{a}k (GR) proposed a new family of the vertex operator algebra (VOA) as the symmetry appearing at an intersection of five-branes to which they refer as $Y$ algebra. Proch\'azka and Rap\v{c}\'{a}k, then proposed to interpret $Y$ algebra as a truncation of affine Yangian whose module is directly connected to plane partitions (PP). They also developed GR's idea to generate a new VOA by connecting plane partitions through an infinite leg shared by them and referred it as the web of W-algebra (WoW).  In this paper, we demonstrate that double truncation of PP gives the minimal models of such VOAs. For a single PP, it generates all the minimal model irreducible representations of $W$-algebra. We find that the rule connecting two PPs is more involved than those in the literature when the $U(1)$ charge connecting two PPs is negative. For the simplest nontrivial WoW, $\mathcal{N}=2$ superconformal algebra, we demonstrate that the improved rule precisely reproduces the known character of the minimal models.
\end{abstract}
\vfill
\end{titlepage}

\setcounter{footnote}{0}

\section{Introduction}
After the success of two-dimensional conformal field theory (CFT), the construction of the new chiral algebras and the study of their representation theory such as their minimal models had been an essential issue in string theory during the 80s. The first examples were the superconformal algebra (SCA) and the W-algebras \cite{DiFrancesco:1997nk}.

In the process of proving Alday-Gaiotto-Tachikawa (AGT) duality \cite{Alday2010}, there was a dramatical change in the analysis of the chiral algebra \cite{schiffmann2013cherednik}. Instead of using the module generated by applying the chiral algebra generators, one introduces the orthogonal basis labeled by the Young diagrams, which also distinguish the fixed points in the instanton moduli space of the supersymmetric Yang-Mills theory. Such basis describes a representation of affine Yangian \cite{schiffmann2013cherednik, Maulik:2012wi,feigin2011quantum, Tsymbaliuk:2014}\footnote{This algebra has been studied by many mathematicians and has many other names, such as ``SH$^c$"\cite{schiffmann2013cherednik}, ``quantum continuous $gl(\infty)$" \cite{feigin2011quantum}.}. The equivalence between the $W_n$-algebra with $U(1)$ current and the affine Yangian was essential in the proof of AGT conjecture. Indeed, the affine Yangian is equivalent to $W_{1+\infty}[\mu]$ \cite{Gaberdiel:2012ku,Gaberdiel:2011wb} which contains $W_n$ algebra as its truncation.

Recently, Gaiotto and Rap\v{c}\'{a}k \cite{Gaiotto:2017euk} introduced a vertex operator algebra (VOA) through the intersection of D5, NS5 and $(-1,-1)$ 5-brane and putting various numbers of D3-branes between these 5-branes.  The algebra is called $Y_{L, M, N}[\Psi]$ where $L, M, N$ are the non-negative integers which represent the number of D3-branes and $\Psi\in\mathbb{C}$ is the parameters of the algebra. By the analysis of gauge theory with the interfaces, they showed that the chiral algebra (or VOA) thus constructed can be described as the BRST reduction of various super Lie algebras. At the same time, the use of the trivalent vertex implies that it would be natural to connect them in the form of the Feynman diagram to generate new VOAs. 

Proch\'azka and Rap\v{c}\'{a}k \cite{Prochazka:2017qum} developed this idea further.  They used a realization of $Y_{L, M, N}$ algebra through the affine Yangian.  They used the fact that the plane partition (PP) with three asymptotes written by Young diagrams gave a natural representation of the affine Yangian and identified  $Y_{L, M, N}$ with a degenerate representation with the null state at $(L+1, M+1, N+1)$. Thus they obtained a natural picture of connecting two plane partitions through the identification of one of three asymptotes in two diagrams. They refer to this new construction of the VOA as Web of W-algebra (WoW).  Through such a viewpoint they obtained many examples of new VOAs by using the diagrammatic technique. \footnote{We have to mention that Gaberdiel and his collaborators \cite{Gaberdiel:2017hcn,Gaberdiel:2018nbs} used a similar method to define the $\mathcal{N}=2$ super W-algebra.}

In this paper, we explore graphical realizations of the minimal models of such chiral algebras by PPs. In particular, the focus is on the examples such as $W_n$ algebra and $\mathcal{N}=2$ superconformal algebra where the detail of the minimal models is known.  To start from such examples is nontrivial and useful since the Hilbert space of WoW is written by the combination of the plane partition which is different from the conventional method by the free fields and the screening currents.  It is also helpful to understand the construction \cite{Prochazka:2017qum, Gaberdiel:2017hcn, Gaberdiel:2018nbs} more clearly. In particular, it solves the subtlety in the treatment of the intermediate Young diagrams when the $U(1)$ charge is negative.

We organize this paper as follows.  In section 2, we review some basic features of the affine Yangian and the construction of the VOA in \cite{Gaiotto:2017euk, Prochazka:2017qum}.  In particular, we emphasize that the relevant algebra is a truncated version of the affine Yangian where one imposes the PPs to have a ``pit" \cite{bershtein2018plane}. In section 3, we demonstrate the explicit construction of WoW by the free fermion and the bosonic ghost.  While it is straightforward, it is illuminative to understand how the intermediate Young diagrams appear explicitly in the  PPs. In section 4, we introduce double truncation of the affine Yangian, which happens when we impose an additional constraint. We will relate it to the minimal models in the later sections. In section 5, we review the PP with nonvanishing asymptotic Young diagrams, which describes the nontrivial representations of the affine Yangian. We use the formula for the conformal dimension and $U(1)$ charge as a hint, and we propose a way to interpret asymptotic Young diagram with negative rows, which will be necessary for WoW.  In section 6, we give an explicit PP realization of  $W_n$ algebra minimal models.  We show that the double truncation with appropriate asymptotes as a set of Young diagrams satisfying $n$-Burge condition \cite{Burge:1993, Belavin:2015ria,Alkalaev:2014sma} which precisely characterizes the minimal model. In section 7, we describe $\mathcal{N}=2$ SCA by two PPs and give a formula for the conformal dimension and $U(1)$ charge \cite{Prochazka:2017qum, Gaberdiel:2017hcn}. In section 8, we study the minimal model by the double truncation and the modified treatment of asymptotic Young diagrams. We show that the conformal dimension, $U(1)$ charge, and the character coincide with the literature \cite{Ravanini:1987yg, Matsuo:1986cj,Dobrev:1986hq,Kiritsis:1986rv}. From these examples, it is convincing that the double truncation of the connected PPs describes of the minimal models in WoW.

\section{A brief review of affine Yangian and Web of W}
\label{sec:aYWoW}

\subsection{Affine Yangian $Y(\widehat{\mathfrak{gl}(1)})$: definition and plane partition representation}
\label{subsec:defrep}
We start from the definition of the affine Yangian $Y(\widehat{\mathfrak{gl}(1)})$ by Drinfeld currents
\begin{equation}
e(u)=\sum_{j=0}^{\infty}\frac{e_j}{u^{j+1}},\qquad
f(u)=\sum_{j=0}^{\infty}\frac{f_j}{u^{j+1}},\qquad
\psi(u)=1+\sigma_3\sum_{j=0}^{\infty}\frac{\psi_j}{u^{j+1}}.
\end{equation}
The parameter $u$ in the currents is the spectral parameter which appears in the integrable model. 
%
We follow the notation in \cite{Prochazka:2015deb} where
the algebra is  parametrized by three parameters $h_1, h_2, h_3\in \mathbb{C}$ with a constraint,
\begin{equation}
\label{eq:sumh}
h_1+h_2+h_3=0 .
\end{equation}
It is convenient to introduce 
\begin{equation}
\sigma_2=h_1h_2+h_2h_3+h_3h_1,\qquad \sigma_3=h_1h_2h_3.
\end{equation}
The defining relations among the Drinfeld currents are 
\begin{equation}
\begin{split}
e(u)e(v)&\sim\varphi(u-v)e(v)e(u),\qquad
f(u)f(v)\sim\varphi(v-u)f(v)f(u),\\
\psi(u)e(v)&\sim\varphi(u-v)e(v)\psi(u),\qquad
\psi(u)f(v)\sim\varphi(v-u)f(v)\psi(u),
\end{split}
\end{equation}
\begin{equation}
[\psi_i,\psi_j]=0,\quad[e_i,f_j]=\psi_{i+j},
\end{equation}
\begin{equation}
\begin{split}
&[\psi_0,e_j]=0,\qquad[\psi_1,e_j]=0,\qquad[\psi_2,e_j]=2e_j,\\
&[\psi_0,f_j]=0,\qquad[\psi_1,f_j]=0,\qquad[\psi_2,f_j]=-2f_j
\end{split}
\end{equation}
where $\varphi(u)$ is the structure function,
\begin{equation}
\varphi(u)=\frac{(u+h_1)(u+h_2)(u+h_3)}{(u-h_1)(u-h_2)(u-h_3)}.
\end{equation}
We note that "$\sim$" implies both sides are equal up to regular terms at $u=0$ or $v=0$.  $\psi_0$ is the center of the algebra.
We also impose Serre relations 
\begin{equation}
{\rm Sym}_{(j_1,j_2,j_3)}[e_{j_1},[e_{j_2},e_{j_3+1}]]=0,\quad{\rm Sym}_{(j_1,j_2,j_3)}[f_{j_1},[f_{j_2},f_{j_3+1}]]=0.
\end{equation}
Proch\'azka \cite{Prochazka:2014gqa}  introduced new parameters $\lambda_i\in \mathbb{C}$ ($i=1,2,3$) which is convenient to describe the null states associated with the plane partition. They are related to $h_i$ by \begin{eqnarray}
\lambda_i=-\frac{\psi_0\sigma_3}{h_i}
\end{eqnarray}
The relation (\ref{eq:sumh}) is replaced by
\ba\label{eq:sumhlam}
\sum_{i=1}^3 \lambda_i^{-1}=0.
\ea

$W_{1+\infty}$-algebra \cite{Gaberdiel:2012ku,Gaberdiel:2011wb} is a $W_n$-algebra in $n\rightarrow \infty$ limit with an extra decoupled $U(1)$ current. 
$W_{1+\infty}$ looks very different from the affine Yangian since the parameter $z$ in the current is a coordinate of world sheet. The algebra contains two parameters, $c$ and $x$, where $c$ is the central charge of Virasoro algebra and $x$ is the parameter which describes the OPE coefficients of higher currents.
While $W_{1+\infty}$ is described by the currents which are familiar in string theory, it may not be so convenient to see the relation with Nekrasov instanton partition function \cite{Nekrasov:2002qd} or the topological string amplitude \cite{Aganagic:2003db, Iqbal:2007ii, Awata:2008ed}. For such purpose, it is better to use the equivalent affine Yangian. The parameters of $W_{1+\infty}$ and the affine Yangian are related by \cite{Prochazka:2014gqa},
\ba
\label{eq:Winftyparameter}
c=1+\prod_{i=1}^3 (\lambda_i-1), \quad
x^2=144(c+1) \prod_{i=1}^3 (\lambda_i-2)(\lambda_i-3)^{-1}\,,
\ea
where $c$ is the central charge of Virasoro algebra and $x^2=\frac{C_{44}^0(C_{33}^4)^2}{(C_{33}^0)^2}$ is a parameter defined in terms of OPE coefficients among primary fields.

In the following, we use both notations, $h_i$ and $\lambda_i$, depending on the context. The use of $\lambda$ has an advantage that one can represent the location of the null state smartly.  For instance, when one of the $\lambda_i$ is a positive integer $N$, the $W_{1+\infty}$ algebra is reduced to $W_N$ algebra with $U(1)$ current. On the other hand, the redundancy in $h_i$ is more useful in the definition of the representation.


\subsection{Representation by plane partition}
There are two types of the representations of the affine Yangian. For instances, there is a free boson realization which is natural in the viewpoint of $W$-algebra. The other representation uses the basis labeled by $n$-tupple Young diagrams, which is useful in the correspondence with the Nekrasov partition functions. We refer the first (resp. second) realization as the horizontal (resp. vertical) realization. The equivalence between the two was essential to prove AGT conjecture \cite{schiffmann2013cherednik,Alday2010,Wyllard2009}.

In the following, we consider representation by a plane partition \cite{Tsymbaliuk:2014,Prochazka:2015deb} which is a natural generalization of the vertical representation. We introduce a set of basis with a label of a plane partition $\Lambda$, and it spans the Hilbert space of the algebra.
The operator $\psi_i$ is diagonal with respect to $|\lambda\rangle$ and $e_i$($f_i$) play a role of adding (removing) a box to $\Lambda$:
\begin{eqnarray}
\psi(u)\ket{\Lambda}&=&\psi_{\Lambda}(u)\ket{\Lambda},\\
\label{eq:addbox}
e(u)\ket{\Lambda}&=&\sum_{\raisebox{1.7pt}{\fbox{}}\in\Lambda^+}\frac{1}{u-q-h_{{\fbox{}}}}\sqrt{-\frac{1}{\sigma_3}{\rm res}_{u\to q+h_{\fbox{}}}\psi_{\Lambda}(u)}\ket{\Lambda+\raisebox{3.5pt}{\fbox{}}},\\
\label{eq:removebox}
f(u)\ket{\Lambda}&=&\sum_{\raisebox{1.7pt}{\fbox{}}\in\Lambda^-}\frac{1}{u-q-h_{{\fbox{}}}}\sqrt{-\frac{1}{\sigma_3}{\rm res}_{u\to q+h_{\fbox{}}}\psi_{\Lambda-\raisebox{2.2pt}{\fbox{}}}(u)}\ket{\Lambda-\raisebox{3.5pt}{\fbox{}}},
\end{eqnarray}
where 
\begin{equation}
\label{eq:psieigenvalue}
\begin{split}
&\psi_{\Lambda}(u)=\psi_0(u-q)\prod_{\raisebox{1.5pt}{\fbox{}}\in\Lambda}\varphi(u-q-h_{\fbox{}}),\\
&\psi_0(u)=\frac{u+\psi_0\sigma_3}{u}.
\end{split}
\end{equation}
Here, $\Lambda^{\pm}$ are the places where we can add (or remove) boxes so that the shape of plane partition is consistent.
We introduce a coordinate for each box in the plane partition. We assign $(1,1,1)$ to the origin of the partition and $(x_1,x_2,x_3)\in (\mathbb{Z}_{>0})^{\otimes 3}$ for a general box. We assign
\begin{equation}
\label{eq:assignedvalue}
h_{\fbox{}}=h_1x_1+h_2x_2+h_3x_3
\end{equation}
to the box located at $(x_1,x_2,x_3)$. We introduce an extra parameter ``$q$" to represent the shift of the spectral parameter. It gives an automorphism of affine Yangian. While it does not change the structure of representation, it represents the charge of $U(1)$ factor. In the following, we use the representation with $q=0$ when we do not mention it explicitly.
\footnote{
We may cancel the redundancy of $h_1,h_2,h_3$ and the center $\psi_0$ by scaling $u$.}

As we already mentioned, Proch\'azka's parametrization has an advantage that the reduction of the representation becomes manifest. 
In the example we mentioned (one of $\lambda_i$, say $\lambda_3$, becomes positive integer), the basis $|\Lambda\rangle$ becomes null when it contains a box at $(1, 1,N+1)$. 
With such condition, the height of the plane partition $\Lambda$ for the nonvanishing states is not greater than $N$. 
One may decompose the plane partition layer by layer into $N$-tuple Young diagrams $Y_1,\cdots, Y_N$ with the condition $Y_1\succeq\cdots \succeq Y_N$.  They give a representation space of $W_N$ algebra with an extra $U(1)$ factor. 

In general, when $\lambda_i$ satisfies the extra condition
\ba\label{DegLMN}
\frac{L}{\lambda_1}+\frac{M}{\lambda_2}+\frac{N}{\lambda_3}=1,
\ea
the basis $|\Lambda\rangle$ which contains a box with a coordinate $(L+1, M+1, N+1)$ becomes null.
Proch\'azka and Rap\v{c}\'{a}k \cite{Prochazka:2017qum} claimed that the affine Yangian whose parameter is constrained by this condition is equivalent to the vertex operator algebra $Y_{L,M,N}[\Psi]$ in \cite{Gaiotto:2017euk}. 

We can derive these null state conditions from (\ref{eq:addbox}) and (\ref{eq:removebox}). The equation
\begin{equation}
{\rm res}_{u\to h_{\fbox{}}}\psi_{\Lambda}(u)=0\qquad(\raisebox{2.5pt}{\fbox{}}\in\Lambda^+).
\end{equation}
implies that the application of the Drinfeld current $e(z)$ cannot generate the state $\ket{\Lambda+\raisebox{3.5pt}{\fbox{}}}$ since the coefficient attached to the basis vanishes. Because $\psi_{\Lambda}(u)$ contains a factor $\frac{u+\psi_0\sigma_3}{u-h_{\fbox{}}}$ for any $\raisebox{2.3pt}{\fbox{}}\in\Lambda^+$ in generic parameters, this happens if 
\begin{equation}
\label{eq:truncation}
\psi_0\sigma_3=-h_{\fbox{}},
\end{equation}
or equivalently
\begin{equation}
\label{eq:truncation2}
\sum_{i=1}^3\frac{x_i}{\lambda_i}=1\,.
\end{equation}
We may interpret it as the null state condition.

We note that the condition (\ref{DegLMN}) has a shift symmetry, 
\begin{eqnarray}
L\rightarrow L+k,\quad
M\rightarrow M+k,\quad
N\rightarrow N+k.
\end{eqnarray}
for $k\in \mathbb{Z}$ due to (\ref{eq:sumhlam}). It allows the redefinition the location of the pit such that the smallest element is one and others are greater or equal to one.
The character of the plane partitions with a pit was derived in \cite{bershtein2018plane}.

The plane partition may have non-trivial asymptotes written in the form of three Young diagrams $\mu_1, \mu_2, \mu_3$ for each axis $x_1, x_2, x_3$.  The partition function for nontrivial plane partition  is given by the topological vertex \cite{Okounkov:2003sp},
\ba
\chi[q]=C_{\mu_1,\mu_2,\mu_3}(q) \prod_{n=1}^\infty (1-q^n)^{-n}\,.
\ea
When there is a pit at $(L+1, M+1, N+1)$, the asymptotic Young diagrams have a similar pit, for example, $\mu_1$ has a pit at $(M+1, N+1)$.

\subsection{Y-algebra and WoW}
\label{subsec:Y-algebra}
Gaiotto and Rap\v{c}\'{a}k \cite{Gaiotto:2017euk} constructed a new VOA by glueing 5-brane junction with some D3 branes set as  follows:
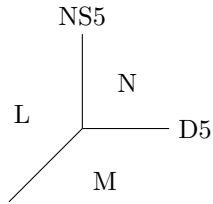
\begin{center}
\begin{tikzpicture}
\footnotesize
\node (D5) at (1.5,0) {D5 };
\node (NS5) at (0,1.5) {NS5 };
\node (dionic5) at (-1.1,-1.1) { };
\draw (0,0)--(D5);
\draw (0,0)--(NS5);
\draw (0,0)--(dionic5);
\node (D3L) at (-0.8,0.2) {L};
\node (D3M) at (0.3,-0.7) {M};
\node (D3N) at (0.6,0.6) {N};
\end{tikzpicture}
\end{center}
Here, $L,M$ and $N$ indicate the number of D3 branes. A twisted $\mathcal{N}=4$ SYM with $U(L)$, $U(M)$ or $U(N)$ gauge group lives on each of D3 branes and Chern-Simons theory with $U(N|L), U(M|L)$ or $U(M|N)$ gauge group does on the interfaces between two of them. The coupling $\Psi$ of SYM is related to Chern-Simons level $k$ as
\begin{equation}
\Psi=k+h \ ,
\end{equation}
where $h$ is the dual-Coxeter  number.
In this setup,  the VOA called Y-algebra $Y_{L,M,N}[\Psi]$ arises at the corner in the above diagram. 
It is defined as follows. Let's focus on NS5 brane and $(-1,-1)$ brane. We can see the corner in the above diagram as  the boundary where $U(N|L)$ and $U(M|L)$ Chern-Simons theories meet. As is well known, 2d chiral algebra arises there, which depends on the boundary condition. In the setup in \cite{Gaiotto:2017euk}, the gauge group is partially preserved as $U(N|L)\to U(M|L)$ if $N>M$. As a consequence, $\widehat{U}(N|L)_{\Psi}$ affine-Kac-Moody algebra is reduced to $\mathcal{DS}_{N-M}[\widehat{U}(N|L)_{\Psi}]$. Here, $\mathcal{DS}_{N-M}$ means Drinfeld-Sokolov reduction with the principal $su(2)$ embedding in $(N-M)\times(N-M)$ part in $U(N|L)$. Finally, we obtain the algebra by $u(M|L)$ BRST reduction. As explained in \cite{Gaiotto:2017euk}, it is considered to be equivalent to take coset as follows:
\begin{equation}
\frac{\mathcal{DS}_{N-M}[\widehat{U}(N|L)_{\Psi}]}{\widehat{U}(M|L)_{\Psi-1}}.
\end{equation}
Note that the level of the subalgebra $\widehat{U}(M|L)$  in $\widehat{U}(N|L)$ is changed due to the contribution from the triangular constituent necessary to make it BRST-closed.
One can consider the $N<M$ case in the same way. It is true of the $N=M$ case except for the existence of the matter $\mathcal{S}^{N|L}$ at the corner, where  $\mathcal{S}^{N|L}$ means $N$ sympletic bosons  and $L$ free fermions.

The vacuum character of $Y_{L,M,N}[\Psi]$ is same as that of a plane partition with a pit at ($L+1, M+1, N+1$), from which Proch\'azka and Rap\v{c}\'{a}k \cite{Prochazka:2017qum} claimed that it is equivalent to the truncation of the affine Yangian. 
The degenerate modules of Y-algebra arise by introducing line operators on the interfaces labeled by the weight of $U(N|L), U(M|L)$ or $U(M|N)$. One may relate them to the asymptotic Young diagrams.

By gluing the plane partitions through the asymptotic Young diagram, one can obtain the Hilbert space of an extended algebra. For example, the following diagram represents $\mathcal{N}=2$ super Virasoro algebra $\otimes \ U(1)$ current:

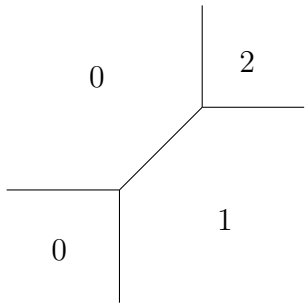
\begin{center}
\begin{tikzpicture}
\node (D5) at (1.5,0) { };
\node (NS5) at (0,1.5) { };
\node (dionic5) at (-1.1,-1.1) { };
\draw (0,0)--(D5);
\draw (0,0)--(NS5);
\draw (0,0)--(-1.1,-1.1);
\draw (-2.6,-1.1)--(-1.1,-1.1);
\draw (-1.1,-2.6)--(-1.1,-1.1);
\node (D3L) at (-1.4,0.4) {0};
\node (D3M) at (0.3,-1.5) {1};
\node (D3N) at (0.6,0.6) {2};
\node (period) at (-1.9,-1.9) {0};
\end{tikzpicture}
\end{center}
One may explain it by a BRST procedure \cite{Gaiotto:2017euk,Prochazka:2017qum}: the trivial DS reduction $\mathcal{DS}_1$ acts on $U(2)$ affine Kac-Moody algebra at the top corner. Then $U(1)$ free fermion couples to it at the bottom corner and finally one takes the coset of them. It leads to
\begin{equation}
\frac{\widehat{U}(2)_{\Psi}\times {\rm Ff}^{U(1)}}{\widehat{U}(1)_{\Psi}}\ ,
\end{equation}
where Ff implies free fermion. This is the known coset realization of $\mathcal{N}=2$ super Virasoro algebra with an extra $U(1)$ factor.

One may implement such a system by gluing $Y_{0,1,2}$ and $Y_{1,0,0}$ through a shared asymptotic Young diagram. 
Since $Y$-algebra is a generalization of $W$-algebra, we refer to the VOA obtained by gluing them as ``web of W" (WoW).

\section{Plane partition realization of free WoW}
\label{sec:freefield}
Before starting to analyze the minimal models, it will be illuminative to explain a free field realization of WoW\footnote{We mention the reference \cite{Prochazka:2018tlo} as the related work.}. In this case, the affine Yangian reduces to undeformed $W_{1+\infty}$-algebra  \cite{Pope:1989ew,Kac:1993zg,Awata:1994tf}. The reduction was studied, for example, in \cite{Prochazka:2015deb}. The WoW extension of the free system was explored by Gaberdiel et al. \cite{Gaberdiel:2018nbs} in the context of the super W-algebra.  A novelty here is the explicit construction of the truncated plane partitions.

The WoW diagram which we will study is,
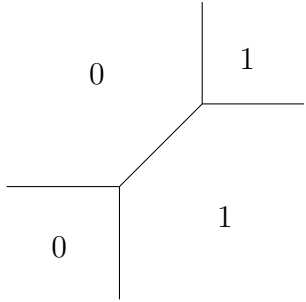
\begin{figure}[H]\label{freeWoW}
\begin{center}
\begin{tikzpicture}
\node (D5) at (1.5,0) { };
\node (NS5) at (0,1.5) { };
\node (dionic5) at (-1.1,-1.1) { };
\draw (0,0)--(D5);
\draw (0,0)--(NS5);
\draw (0,0)--(-1.1,-1.1);
\draw (-2.6,-1.1)--(-1.1,-1.1);
\draw (-1.1,-2.6)--(-1.1,-1.1);
\node (D3L) at (-1.4,0.4) {0};
\node (D3M) at (0.3,-1.5) {1};
\node (D3N) at (0.6,0.6) {1};
\node (period) at (-1.9,-1.9) {0};
\end{tikzpicture}
\end{center}
\end{figure}
We first explain $Y_{1,1,0}$ and $Y_{0,0,1}$\footnote{We note the Y-algebras read off from the diagram are $Y_{0,1,1}$ and $Y_{1,0,0}$. We use the arbitrariness rename the axes to make figures easy to read. We will use the similar trick in the following.} describe a bosonic ghost
and a free fermion respectively.
We follow the free field representations of $W_{1+\infty}$-algebra in \cite{Awata:1994zr} and super $W_{1+\infty}$ in \cite{awata1995quasifinite}. In these references, the authors obtained results which can be straightforwardly related to the explicit form of the truncated plane partition realizations.

\subsection{Bosonic ghost}
We start from the bosonic ghost fields $\beta(z)=\sum_{n\in \mathbb{Z}} \beta_n z^{-n}$, $\gamma(z)=\sum_{n\in \mathbb{Z}} \gamma_n z^{-n-1}$. The commutation relation among the oscillators is given by $\left[\beta_n, \gamma_m\right]=\delta_{n+m,0}$.  The generators of $W_{1+\infty}$ are written as the bilinear combinations $W_{n,m}=\int d\zeta :\beta(\zeta)\zeta^n D_\zeta^m \gamma(\zeta):$ ($n\in\mathbb{Z}$, $m\in \mathbb{Z}_{\geq 0}$ and $D:=\zeta\partial_\zeta$).
We note that the Hilbert space is generated from the vacuum satisfying $\beta_n|0\rangle =0$ ($n\geq 0$) and $\gamma_n|0\rangle =0$ ($n>0$) by applying such bilinear operators. The basis takes the form
\ba
\beta_{-n_1}\cdots \beta_{-n_g} \gamma_{-m_1} \cdots \gamma_{-m_g}|0\rangle
\ea
where $n_1\geq \cdots \geq n_g\geq 1$ and $m_1\geq m_2\geq \cdots \geq m_g\geq 0$. We note that the ghost number should vanish.

One may construct two Young diagrams $(n_1,\cdots, n_g)$ and $(m_1,\cdots, m_g)$ and attach the second piece to the first one perpendicularly. 
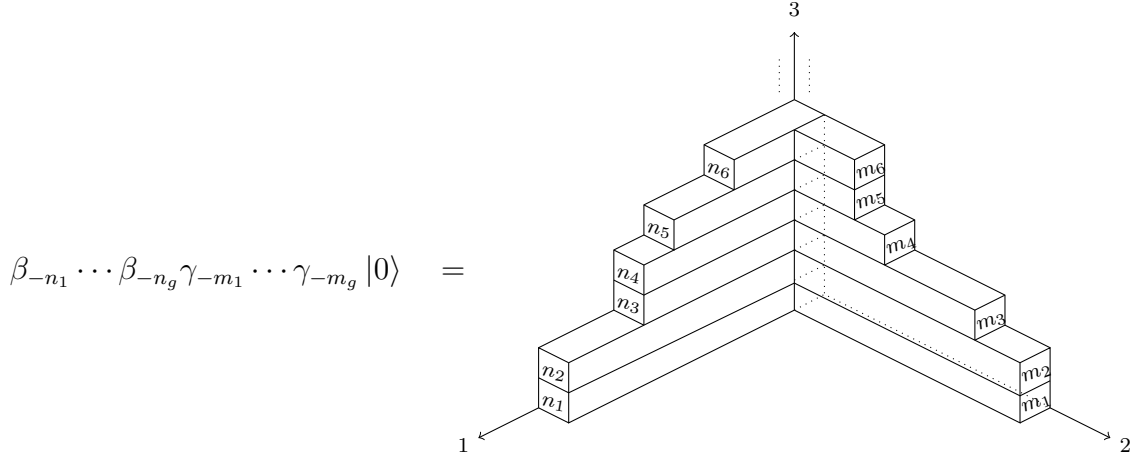
\begin{figure}[H]
	\begin{center}
		\begin{tikzpicture}
		\node (white) at (-7.4,0.5) {${\displaystyle\beta_{-n_1}\cdots\beta_{-n_g}\gamma_{-m_1}\cdots\gamma_{-m_g}\ket{0}\quad=}$};
		\scriptsize
		\draw (0,0)--(0,2.4);
		\draw (0,0)--(3,-1.5)--(3,-0.7)--(0,0.8);
		\draw (2.4,-0.4)--(2.4,-0)--(0,1.2);
		\draw (1.2,0.6)--(1.2,1)--(0,1.6);
		\draw (0.8,1.2)--(0.8,2)--(0,2.4);
		\draw  (0,2)--(0.8,1.6)--(1.2,1.8);
		
		\draw (3,-1.5)--(3.4,-1.3)--(3.4,-0.5)--(3,-0.7);
		\draw (2.4,-0.4)--(2.8,-0.2);
		\draw (2.4,0)--(2.8,0.2);
		\draw (1.2,0.6)--(1.6,0.8);
		\draw (1.2,1)--(1.6,1.2);
		\draw (0.8,1.2)--(1.2,1.4);
		\draw (0.8,2)--(1.2,2.2);
		\draw (0,2.4)--(0.4,2.6);
		
		\draw (3.4,-0.5)--(2.8,-0.2)--(2.8,0.2)--(1.6,0.8)--(1.6,1.2)--(1.2,1.4)--(1.2,2.2)--(0.4,2.6);
		
		\draw (0,0)--(-3,-1.5)--(-3,-0.7)--(0,0.8);
		\draw (-0.8,1.6)--(-0.8,2)--(0,2.4);
		
		\draw (-3,-1.5)--(-3.4,-1.3)--(-3.4,-0.5)--(-3,-0.7);
		\draw (-2,-0.2)--(-2.4,0);
		\draw (-0.8,1.6)--(-1.2,1.8);
		\draw (-0.8,2)--(-1.2,2.2);
		\draw (-3.4,-0.5)--(-2.4,0)--(-2.4,0.8)--(-2,1)--(-2,1.4)--(-1.2,1.8)--(-1.2,2.2)--(0,2.8)--(0.4,2.6);
		\draw (-2,1.4)--(-1.6,1.2)--(-1.6,0.8)--(-2,1);
		\draw (-1.6,1.2)--(0,2);
		\draw (-2,0.6)--(-2,-0.2);
		
		\draw (-3.4,-0.9)--(-3,-1.1)--(0,0.36);
		\draw (-2.4,0.4)--(-2,0.2)--(0,1.2);
		\draw (-2.4,0.8)--(-2,0.6)--(0,1.6);
		\draw (0,0.36)--(3,-1.14)--(3.4,-0.94);
		
		\draw[->] (0,2.8)--(0,3.7);
		\draw[->] (-3.4,-1.3)--(-4.2,-1.7);
		\draw[->] (3.4,-1.3)--(4.2,-1.7);
		
		\node (1) at (-4.4,-1.8) {1};
		\node (2) at (4.4,-1.8) {2};
		\node (3) at (0,4) {3};
		
		\draw[dotted] (0.4,2.6)--(0.4,0.2);
		\draw[dotted] (0,0)--(0.4,0.21)--(3.4,-1.29);
		\draw[dotted] (0,0.8)--(0.4,1);
		\draw[dotted] (0,1.2)--(0.4,1.4);
		\draw[dotted] (0,1.6)--(0.4,1.8);
		\draw[dotted] (0,2)--(0.4,2.2);
		\draw[dotted] (0,0.4)--(0.4,0.6);
		
		\node (n1) at (-3.2,-1.25) {\rotatebox{-10}{$n_1$}};
		\node (n2) at (-3.2,-0.85) {\rotatebox{-10}{$n_2$}};
		\node (n3) at (-2.2,0.05) {\rotatebox{-10}{$n_3$}};
		\node (n4) at (-2.2,0.45) {\rotatebox{-10}{$n_4$}};
		\node (n5) at (-1.8,1.05) {\rotatebox{-10}{$n_5$}};
		\node (n6) at (-1,1.85) {\rotatebox{-10}{$n_6$}};
		
		\node (m1) at (3.23,-1.25) {\rotatebox{20}{$m_1$}};
		\node (m2) at (3.23,-0.85) {\rotatebox{20}{$m_2$}};
		\node (m3) at (2.63,-0.15) {\rotatebox{20}{$m_3$}};
		\node (m4) at (1.43,0.85) {\rotatebox{20}{$m_4$}};
		\node (m5) at (1.03,1.45) {\rotatebox{20}{$m_5$}};
		\node (m6) at (1.03,1.85) {\rotatebox{20}{$m_6$}};

		
		
		
		\draw[dotted] (0.2,2.9)--(0.2,3.4);
		\draw[dotted] (-0.2,2.9)--(-0.2,3.4);
		
		
		
		
		
		
		
		\end{tikzpicture}
	\end{center}
	\caption{The relation between Hilbert space of $\beta\gamma$ ghost and plane partition with a pit at (2,2,1). It can be decomposed into two Young diagrams. The left (right) one  corresponds to the Hilbert space of $\beta$ ($\gamma$) ghost. The number written in each row means its length.}
\end{figure}
It defines a plane partition with a pit at $(2,2,1)$ and thus describes a module of $Y_{1,1,0}$. One may obtain the parameters of the affine Yangian by imposing a condition $\lambda_1^{-1}+\lambda_2^{-1}=1$ such that we have a pit at $(2,2,1)$. One may solve it with (\ref{eq:sumhlam}) by $\lambda_3=-1$ and $\lambda_2=\frac{\lambda_1}{\lambda_1-1}$. The formula (\ref{eq:Winftyparameter}) implies $c=-1$ for any $\lambda_1$. This is the central charge for the bosonic ghost.  We note that the affine Yangian reduces to $W_{1+\infty}$ without the deformation parameter in the self-dual limit (one of $\lambda_i$ is infinite). This condition is met if we set $\lambda_1=1$ (or $\lambda_2=1$).

The partition function becomes,
\begin{eqnarray}
\chi(q)=\sum_{g=0}^\infty\sum_{n_1\geq\cdots\geq n_g\geq 1}\sum_{m_1\geq\cdots \geq m_g\geq 0}q^{\sum_i (n_i+m_i)}
=\sum_{n=0}^\infty \frac{q^n}{((q;q)_n)^2},\qquad (a;q)_n=\prod_{m=0}^{n-1}(1-aq^m)\,.
\end{eqnarray}
When there are $N$-pairs of bosonic ghosts, we obtain a similar plane partition where the location of the pit moves to $(N+1, N+1, 1)$ which corresponds to $Y_{N,N,0}$ ($\lambda_1=-\lambda_3=N, \lambda_2=\infty$, $c=-N$). In \cite{Awata:1994tf}, one may find the explicit form of the partition function for such generalized case.

\subsection{Free fermion}
Similarly the fermionic Hilbert space described by $\{ b_n, c_m\}=\delta_{n+m,0}$ is spanned by 
\ba
b_{-n_1}\cdots b_{-n_g} c_{-m_1} \cdots c_{-m_g}|0\rangle
\ea
where $n_1>\cdots > n_g\geq 1$ and $m_1> m_2> \cdots > m_g\geq 0$. 
Again the state is constructed out of bilinear of fermionic ghosts, and the ghost number is zero.  In this case, one can associated a Young diagram by combining hooks $(n_l, 1^{m_l})$ in the order $l=1,2,\cdots, g$.

\begin{figure}[H]
	\begin{center}
		\begin{tikzpicture}
		
		\node (white) at (-3.3,-1.5) {${\displaystyle b_{-n_1}\cdots b_{-n_g}c_{-m_1}\cdots c_{-m_g}\ket{0}\quad=}$};
		\scriptsize
		\draw (0,0) rectangle (4.5,-0.4);
		\draw (0,-0.4) rectangle (0.4,-3.5);
		\draw (0.4,-0.4) rectangle (4.0,-0.8);
		\draw (0.4,-0.8) rectangle (0.8,-3.0);
		\draw[dotted] (0.9,-0.9)--(1.1,-1.1);
		\draw (1.2,-1.2) rectangle (3.5,-1.6);
		\draw (1.2,-1.6) rectangle (1.65,-2.4);
		\node (white) at (2.5,-0.25) {$n_1$};
		\node (white) at (2.5,-0.65) {$n_2$};
		\node (white) at (2.5,-1.45) {$n_g$};
		\node (white) at (0.22,-2.5) {$m_1$};
		\node (white) at (0.62,-2.3) {$m_2$};
		\node (white) at (1.45,-2.0) {$m_g$};
		
		\end{tikzpicture}
	\end{center}
	\caption{The relation between Hilbert space of $bc$ ghost and Young diagram (plane partition with a pit at (1,1,2)). The number written in each row or column means its length.}
\end{figure}
The partition function becomes
\begin{eqnarray}
\chi(q)=\sum_{g=0}^\infty\sum_{n_1>\cdots> n_g\geq 1}\sum_{m_1>\cdots > m_g\geq 0}q^{\sum_i (n_i+m_i)}
=\sum_{n=0}^\infty \frac{q^{n^2/2}}{((q;q)_n)^2}=\prod_{n=1}^\infty (1-q^n)^{-1}\,.
\end{eqnarray}
We note that the formula in the third term resembles that of the bosonic ghost.

In this case, the plane partition is truncated to a single Young diagram. The location of the pit is $(1,1,2)$. The condition to have a pit there implies $\lambda_3=1$. One may solve (\ref{eq:sumhlam}) by $\lambda_2=-\frac{\lambda_1}{\lambda_1+1}$. The central charge (\ref{eq:Winftyparameter}) gives $c=1$ for any $\lambda_1$. This is the central charge of free fermion. The self-duality condition is met if $\lambda_1=-1$ and $\lambda_2=\infty$. If there are $N$ fermions, the pit moves to $(1,1,N+1)$.  There is the explicit form of the character for such a diagram \cite{Awata:1994tf} for the self-dual case.

\subsection{WoW: $\mathcal{N}=2$ superconformal algebra}
Finally, we combine $bc$ and $\beta\gamma$ systems. This is the classical representation of $\mathcal{N}=2$ superconformal algebra\footnote{We note that the same web diagram was studied in \cite{Prochazka:2017qum}, where the authors obtained $U(1|1)$ current algebra from DS reduction. 
$U(1|1)$ currents consist of the bilinear combinations of the ghost fields where their conformal dimensions are adjusted so that the dimension of the currents become one. We note that one may shift the dimension of bosonic (or fermionic) ghost fields by $\frac{1}{2}$ to realize $\mathcal{N}=2$ SCA instead of $U(1|1)$. Strictly speaking, the latter interpretation does not respect the implication of the diagram as the local operators inserted at the corner. We thank M.Rap\v{c}\'{a}k for the explanation.} \cite{Friedan:1985ge}. Besides the bilinear operator of $bc$ and $\beta\gamma$,
we include the extra generators which are expressed by the bilinear form
$b\gamma$ and $\beta \partial c$.  If the algebra is extended by these generators,
the bosonic and fermionic ghost numbers do not separately vanish.
The basis of the Hilbert space becomes,
\ba\label{Hilbfree3}
\beta_{-n_1}\cdots \beta_{-n_{g_1}} \gamma_{-m_1} \cdots \gamma_{-m_{g_2}}
b_{-\bar n_1}\cdots b_{-\bar n_{g_3}} c_{-\bar m_1} \cdots c_{-\bar m_{g_4}}|0\rangle
\ea
with $g_1+g_3=g_2+g_4$ but $g_1\neq g_2$, $g_3\neq g_4$ in general.
Suppose $g_1-g_2=h>0$, one may fill the unbalanced rows by an infinite leg with hight $h$ on the right.  On the other hand, with $g_4-g_3=h$, one may similarly fill the other wing of the Young diagram with an infinite leg similarly.

\begin{figure}[H]
	\begin{center}
		\begin{tikzpicture}
		\scriptsize
		\draw (0,0)--(0,2.4);
		\draw (1.2,0.6)--(1.2,1)--(0,1.6);
		\draw (0.8,1.2)--(0.8,2)--(0,2.4);
		\draw  (0,2)--(0.8,1.6)--(1.2,1.8);
		
		\draw (1.2,0.6)--(1.6,0.8);
		\draw (1.2,1)--(1.6,1.2);
		\draw (0.8,1.2)--(1.2,1.4);
		\draw (0.8,2)--(1.2,2.2);
		\draw (0,2.4)--(0.4,2.6);
		
		\draw (1.6,0.8)--(1.6,1.2)--(1.2,1.4)--(1.2,2.2)--(0.4,2.6);
		
		\draw (0,0)--(-3,-1.5)--(-3,-0.7)--(0,0.8);
		\draw (-0.8,1.6)--(-0.8,2)--(0,2.4);
		
		\draw (-3,-1.5)--(-3.4,-1.3)--(-3.4,-0.5)--(-3,-0.7);
		\draw (-2,-0.2)--(-2.4,0);
		\draw (-0.8,1.6)--(-1.2,1.8);
		\draw (-0.8,2)--(-1.2,2.2);
		\draw (-3.4,-0.5)--(-2.4,0)--(-2.4,0.8)--(-2,1)--(-2,1.4)--(-1.2,1.8)--(-1.2,2.2)--(0,2.8)--(0.4,2.6);
		\draw (-2,1.4)--(-1.6,1.2)--(-1.6,0.8)--(-2,1);
		\draw (-1.6,1.2)--(0,2);
		\draw (-2,0.6)--(-2,-0.2);
		
		\draw (-3.4,-0.9)--(-3,-1.1)--(0,0.36);
		\draw (-2.4,0.4)--(-2,0.2)--(0,1.2);
		\draw (-2.4,0.8)--(-2,0.6)--(0,1.6);
		
		\draw[->] (0,2.8)--(0,3.7);
		\draw[->] (-3.4,-1.3)--(-4.2,-1.7);
		\draw[->] (3.4,-1.3)--(4.2,-1.7);
		
		\node (1) at (-4.4,-1.8) {1};
		\node (2) at (4.4,-1.8) {2};
		\node (3) at (0,4) {3};
		
		\draw[dotted] (0.4,2.6)--(0.4,0.2);
		\draw[dotted] (0,0)--(0.4,0.21)--(3.4,-1.29);
		\draw[dotted] (0,0.8)--(0.4,1);
		\draw[dotted] (0,1.2)--(0.4,1.4);
		\draw[dotted] (0,1.6)--(0.4,1.8);
		\draw[dotted] (0,2)--(0.4,2.2);
		\draw[dotted] (0,0.4)--(0.4,0.6);
		
		\node (n1) at (-3.2,-1.25) {\rotatebox{-10}{$n_1$}};
		\node (n2) at (-3.2,-0.85) {\rotatebox{-10}{$n_2$}};
		\node (n3) at (-2.2,0.05) {\rotatebox{-10}{$n_3$}};
		\node (n4) at (-2.2,0.45) {\rotatebox{-10}{$n_4$}};
		\node (n5) at (-1.8,1.05) {\rotatebox{-10}{$n_5$}};
		\node (n6) at (-1,1.85) {\rotatebox{-10}{$n_6$}};
		
		\node (m4) at (1.43,0.85) {\rotatebox{20}{$m_1$}};
		\node (m5) at (1.03,1.45) {\rotatebox{20}{$m_2$}};
		\node (m6) at (1.03,1.85) {\rotatebox{20}{$m_3$}};

		\draw[dotted] (0.2,2.9)--(0.2,3.4);
		\draw[dotted] (-0.2,2.9)--(-0.2,3.4);
		
		\draw (0,1.2)--(3.4,-0.5);
		\draw (0,0)--(3.2,-1.6);
		\draw (1.6,0.8)--(3.6,-0.2);
		\fill[pattern=north east lines] (0,1.2)--(3.2,-0.4)--(3,-1.5)--(0,0)--cycle;
		\fill[pattern=north west lines] (1.2,0.6)--(3.2,-0.4)--(3.6,-0.2)--(1.6,0.8)--cycle;
		
		\large
		\node (times) at (4.8,1) {$\otimes$};
		\scriptsize
			\begin{scope}[yshift=3.3cm]
		\fill[pattern=north east lines]  (6,0) rectangle (6.4,-0.4);
		\draw(6,-0.4) rectangle (6.4,-4.7);
		\fill[pattern=north east lines]  (6.4,-0) rectangle (11.4,-0.8);
		\draw (6.4,-0.8) rectangle (6.8,-4.3);
		\draw (7.2,-1.2) rectangle (9.9,-1.6);
		\draw (6.8,-1.2) rectangle (7.2,-4);
		\draw (7.2,-1.6) rectangle (7.6,-3.6);
		\draw[dotted] (7.7,-1.7)--(7.9,-1.9);
		\draw (8,-2) rectangle (9.4,-2.4);
		\draw (8,-2.4) rectangle (8.45,-3);
		\node (white) at (6.2,-3.25) {$\bar{n}_1$};
		\node (white) at (6.6,-3.1) {$\bar{n}_2$};
		\node (white) at (8.5,-1.45) {$\bar{m}_1$};
		\node (white) at (7.0,-2.95) {$\bar{n}_3$};
		\node (white) at (7.4,-2.85) {$\bar{n}_4$};
		\node (white) at (8.25,-2.7) {$\bar{n}_{g_4}$};
		\node (white) at (8.9,-2.25) {$\bar{m}_{g_3}$};
		\draw(6,-0)--(6,-0.4);
		\draw[->](6,0)--(12.1,0);
		\draw(9.9,-1.2)--(11.6,-1.2);
		\fill[pattern=north east lines]  (6.8,-0.8) rectangle (11.4,-1.2);
		\draw[->] (6,-4.7)--(6,-5.1);
		
		\node (1) at (12.4,0) {2};
		\node (2) at (6,-5.4) {1};
		
		\end{scope}

		\end{tikzpicture}
	\end{center}
	\caption{The figure represents the state in the form of  (\ref{Hilbfree3}) with $g_1-g_2=g_4-g_3>0$. The rows with infinite length and height $g_1-g_2$ are inserted. The above case corresponds to $g_1-g_2=3$.} 
\end{figure}
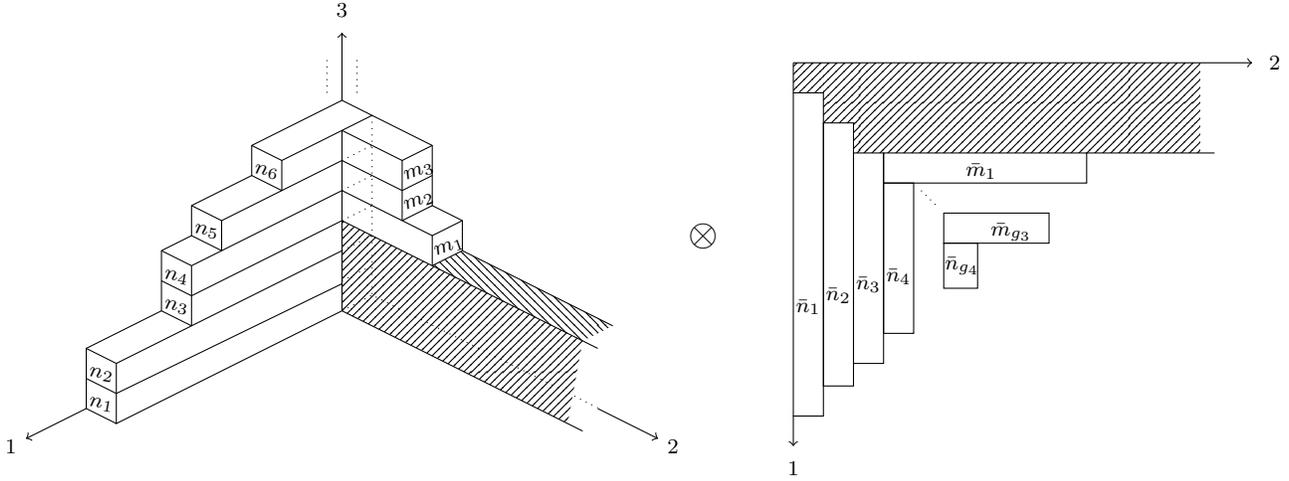

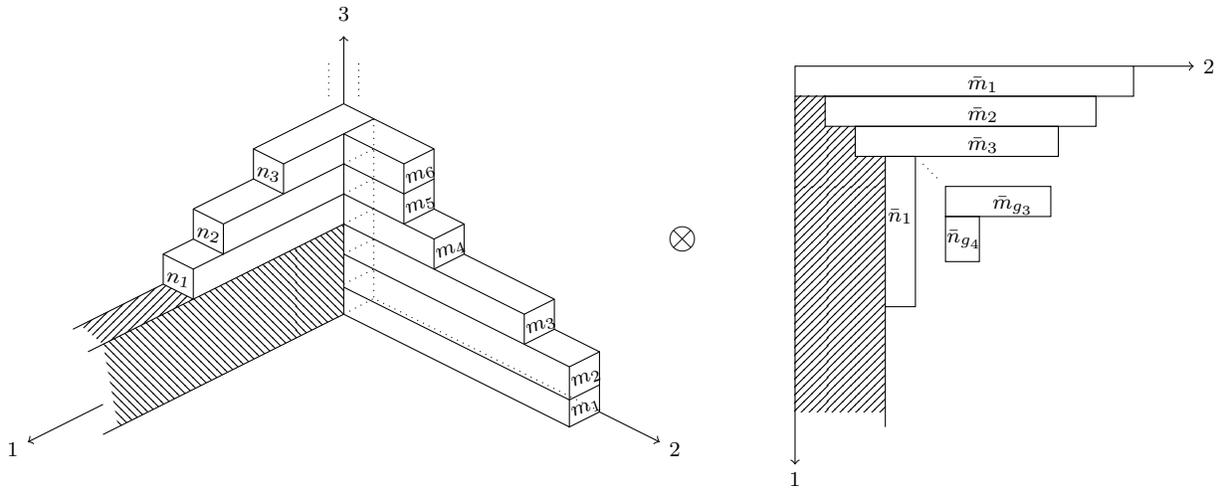
\begin{figure}[H]
	\begin{center}
		\begin{tikzpicture}
		
		\scriptsize
		\draw (0,0)--(0,2.4);
		\draw (0,0)--(3,-1.5)--(3,-0.7)--(0,0.8);
		\draw (2.4,-0.4)--(2.4,-0)--(0,1.2);
		\draw (1.2,0.6)--(1.2,1)--(0,1.6);
		\draw (0.8,1.2)--(0.8,2)--(0,2.4);
		\draw  (0,2)--(0.8,1.6)--(1.2,1.8);
		
		\draw (3,-1.5)--(3.4,-1.3)--(3.4,-0.5)--(3,-0.7);
		\draw (2.4,-0.4)--(2.8,-0.2);
		\draw (2.4,0)--(2.8,0.2);
		\draw (1.2,0.6)--(1.6,0.8);
		\draw (1.2,1)--(1.6,1.2);
		\draw (0.8,1.2)--(1.2,1.4);
		\draw (0.8,2)--(1.2,2.2);
		\draw (0,2.4)--(0.4,2.6);
		
		\draw (3.4,-0.5)--(2.8,-0.2)--(2.8,0.2)--(1.6,0.8)--(1.6,1.2)--(1.2,1.4)--(1.2,2.2)--(0.4,2.6);
		
		\draw (-0.8,1.6)--(-0.8,2)--(0,2.4);
		
		\draw (-0.8,1.6)--(-1.2,1.8);
		\draw (-0.8,2)--(-1.2,2.2);
		\draw (-2.4,0.4)--(-2.4,0.8)--(-2,1)--(-2,1.4)--(-1.2,1.8)--(-1.2,2.2)--(0,2.8)--(0.4,2.6);
		
		\draw (-2,1.4)--(-1.6,1.2)--(-1.6,0.8)--(-2,1);
		\draw (-1.6,1.2)--(0,2);
		\draw (-2,0.6)--(-2,0.2);
		
		\draw (-2.4,0.4)--(-2,0.2)--(0,1.2);
		\draw (-2.4,0.8)--(-2,0.6)--(0,1.6);
		\draw (0,0.36)--(3,-1.14)--(3.4,-0.94);
		
		\draw[->] (0,2.8)--(0,3.7);
		\draw[->] (-3.2,-1.2)--(-4.2,-1.7);
		\draw[->] (3.4,-1.3)--(4.2,-1.7);
		
		\node (1) at (-4.4,-1.8) {1};
		\node (2) at (4.4,-1.8) {2};
		\node (3) at (0,4) {3};
		
		\draw[dotted] (0.4,2.6)--(0.4,0.2);
		\draw[dotted] (0,0)--(0.4,0.21)--(3.4,-1.29);
		\draw[dotted] (0,0.8)--(0.4,1);
		\draw[dotted] (0,1.2)--(0.4,1.4);
		\draw[dotted] (0,1.6)--(0.4,1.8);
		\draw[dotted] (0,2)--(0.4,2.2);
		\draw[dotted] (0,0.4)--(0.4,0.6);
		
		\node (n4) at (-2.2,0.45) {\rotatebox{-10}{$n_1$}};
		\node (n5) at (-1.8,1.05) {\rotatebox{-10}{$n_2$}};
		\node (n6) at (-1,1.85) {\rotatebox{-10}{$n_3$}};
		
		\node (m1) at (3.23,-1.25) {\rotatebox{20}{$m_1$}};
		\node (m2) at (3.23,-0.85) {\rotatebox{20}{$m_2$}};
		\node (m3) at (2.63,-0.15) {\rotatebox{20}{$m_3$}};
		\node (m4) at (1.43,0.85) {\rotatebox{20}{$m_4$}};
		\node (m5) at (1.03,1.45) {\rotatebox{20}{$m_5$}};
		\node (m6) at (1.03,1.85) {\rotatebox{20}{$m_6$}};

		\draw[dotted] (0.2,2.9)--(0.2,3.4);
		\draw[dotted] (-0.2,2.9)--(-0.2,3.4);
		
		\draw (-2,0.2)--(-3.4,-0.5);
		\draw (-2.4,0.4)--(-3.6,-0.2);
		\draw (0,0)--(-3.2,-1.6);
		\fill[pattern=north west lines] (-3.2,-0.4)--(0,1.2)--(0,0)--(-3,-1.5)--cycle;
		\fill[pattern=north east lines] (-2,0.2)--(-2.4,0.4)--(-3.5,-0.15)--(-3.2,-0.4)--cycle;
		
		\large
		\node (times) at (4.5,1) {$\otimes$};
		\scriptsize
		
		\begin{scope}[xshift=6cm,yshift=3.3cm]
		\scriptsize
		\draw (0,0) rectangle (4.5,-0.4);
		\fill[pattern=north east lines] (0,-0.8) rectangle (0.8,-4.6);
		\draw (0.4,-0.4) rectangle (4.0,-0.8);
		\fill[pattern=north east lines] (0,-0.4) rectangle (0.4,-0.8);
		\draw (0.8,-0.8) rectangle (3.5,-1.2);
		\draw (1.2,-1.2) rectangle (1.6,-3.2);
		\fill[pattern=north east lines] (0.8,-1.2) rectangle (1.2,-4.6);
		\draw (1.2,-3.2)--(1.2,-4.8);
		\draw[dotted] (1.7,-1.3)--(1.9,-1.5);
		\draw (2,-1.6) rectangle (3.4,-2.0);
		\draw (2,-2.0) rectangle (2.45,-2.6);
		\node (white) at (2.5,-0.25) {$\bar{m}_1$};
		\node (white) at (2.5,-0.65) {$\bar{m}_2$};
		\node (white) at (2.5,-1.05) {$\bar{m}_3$};
		
		\node (white) at (1.4,-2.0) {$\bar{n}_1$};
		\node (white) at (2.25,-2.3) {$\bar{n}_{g_4}$};
		\node (white) at (2.9,-1.85) {$\bar{m}_{g_3}$};
		
		\draw[->](0,-0.4)--(0,-5.3);
		\draw[->](4.5,0)--(5.3,0);
		\node (1) at (5.5,0) {2};
		\node (2) at (0,-5.5) {1};
		\end{scope}

		\end{tikzpicture}
	\end{center}
	\caption{The figure represents the state in the form of  (\ref{Hilbfree3}) with $g_1-g_2=g_4-g_3<0$. The rows with infinite length and height $g_2-g_1$ are inserted. The above case corresponds to $g_2-g_1=3$.} 
\end{figure}

In this combined system, it is natural to use the connected plane partitions to represent the total Hilbert space. The shared legs have either one row or one column. Depending on the sign of $h$, however, the infinite leg is attached to $x_1$ or $x_2$ direction. It does not fit with the picture presented by the diagram drawn in the first paragraph of this section. We may circumvent this complication by allowing the height or width of the intermediate Young diagram can be negative and suppose the shared asymptotic Young diagram appears in the fixed directions.

The partition function with the summation over the infinite leg becomes
\ba
\chi[q]=\sum_{g_1, g_2, g_3, g_4\geq 0, g_1+g_3=g_2+g_4}
q^{g_1+\frac{1}{2}g_3(g_3+1)+\frac12 g_4(g_4-1)}\prod_{i=1}^4(q;q)^{-1}_{g_i} =\prod_{n=1}^\infty\frac{(1+q^n)^2}{(1-q^n)^2}
\ea
which is the character for the $\mathcal{N}=2$ superconformal algebra (up to a finite factor) in the Ramond sector. One may shift   $\bar n_i, \bar m_i$ in (\ref{Hilbfree3}) to half integers to obtain the character for the NS sector. The discrepancy disappears by noting that the fermionic generators are not the general bilinear combination of $\beta c$ and $b\gamma$  but the restricted ones.  Thus WoW VOA correctly reproduces  the $\mathcal{N}=2$ superconformal algebra with the proper description of the negative $h$.

It is straightforward to generalize the system to an arbitrary number of quartets $(b^{(i)},c^{(i)}, \beta^{(i)}, \gamma^{(i)})$ ($i=1,\cdots, M$) \cite{Gaberdiel:2017hcn,awata1995quasifinite}.
Such system describes the $\mathcal{N}=2$ super $W$-algebra.
The intermediate Young diagrams are the restricted ones whose height or width is limited by $M$.  As in the $M=1$ example, the legs stretch over both $x_1$ and $x_2$ directions.  The explicit analysis is, however, somehow complicated and will be the subject of the future publication.



\section{Double truncation and periodicity in plane partition}
\label{subsec:truncation}
In the following, we consider the special cases where the parameters $\lambda_i$ are subject to two constraints\footnote{
Such relation is trivially satisfied if $L_2=L_1+k$, $M_2=M_1+k$, $N_2=N_1+k$, due to \ref{eq:sumhlam}. We assume the two sets do not satisfy such a relation.} of the form (\ref{DegLMN}),
\begin{eqnarray}\label{eq:double}
\frac{L_1}{\lambda_1}+\frac{M_1}{\lambda_2}+\frac{N_1}{\lambda_3}=1\,,\qquad
\frac{L_2}{\lambda_1}+\frac{M_2}{\lambda_2}+\frac{N_2}{\lambda_3}=1\,.
\end{eqnarray}
With such constraints, we have two pits at $(L_1+1,M_1+1, N_1+1)$ and at $(L_2+1,M_2+1, N_2+1)$ in the plane partition. We need to study carefully which plane partition will be relevant. Such condition was proposed in the case of Lee-Yang singularity in \cite{Prochazka:2015deb}.  In this paper, we focus on the special cases $(L_1,M_1,N_1)=(p,q,0)$ and $(L_2,M_2,N_2)=(0,0,n)$.

When the parameters take such special values where two points satisfies the condition (\ref{eq:truncation2}), we need to re-examine the truncation rule. 
We denote two pits by $\fbox{\footnotesize 1}$ and $\fbox{\footnotesize 2}$ with coordinates $(x_1,y_1,z_1)$ and $(x_2,y_2,z_2)$ respectively and set $h=h_{\fbox{\tiny 1}}=h_{\fbox{\tiny 2}}$. One cannot create a box by the action of $e(u)$ on these pits as long as the both boxes don't belong to $\Lambda^+$.
When both $\fbox{\footnotesize 1}$ and $\fbox{\footnotesize 2}$ belong to $\Lambda^+$, something new happens.
$\psi_\Lambda$ (\ref{eq:psieigenvalue}) contains a factor
\begin{equation}
\psi_{\Lambda}(u)\propto\frac{u+\psi_0\sigma_3}{(u-h)^2}=\frac{1}{u-h} \,.
\end{equation}
Since the residue in (\ref{eq:addbox}) does not vanish,
we can create the states containing either $\fbox{\footnotesize 1}$ or $\fbox{\footnotesize 2}$.  From the viewpoint of the coordinate map $h$, these two boxes cannot be distinguished ($h_{\fbox{\tiny 1}}=h_{\fbox{\tiny 2}}$).  More strongly (\ref{eq:double}) implies that we have a periodicity, $h_{x+L_1, y+M_1, z+N_1}=h_{x+L_2, y+M_2, z+N_2}$. Because of the periodicity, once we fill one of the boxes, say $\fbox{\footnotesize 1}$, one cannot fill the other $\fbox{\footnotesize 2}$ since we cannot add a box twice at the identified position
\footnote{One can also obtain the constraint by considering the null state condition. From (\ref{eq:addbox}) we see that  $e(u)\ket{\Lambda}$ contains the terms proportional to 
$\frac{1}{u-h_{\setlength{\fboxsep}{0.4mm}\fbox{\tiny1}}}\ket{\Lambda+\setlength{\fboxsep}{0.4mm}\fbox{\scriptsize1}\ }+\frac{1}{u-h_{\setlength{\fboxsep}{0.4mm}\fbox{\tiny2}}}\ket{\Lambda+\setlength{\fboxsep}{0.4mm}\fbox{\scriptsize2}\ }=\frac{1}{u-h}\bigl(\ket{\Lambda+\setlength{\fboxsep}{0.4mm}\fbox{\scriptsize1}\ }+\ket{\Lambda+\setlength{\fboxsep}{0.4mm}\fbox{\scriptsize2}\ }\bigr)$. This shows that all modes $e_i$ of $e(u)$ can generate only the one combination $\ket{\Lambda+\setlength{\fboxsep}{0.4mm}\fbox{\scriptsize 1}\ }+\ket{\Lambda+\setlength{\fboxsep}{0.4mm}\fbox{\scriptsize 2}\ }$. Hence, the other one $\ket{\Lambda+\setlength{\fboxsep}{0.4mm}\fbox{\scriptsize 1}\ }-\ket{\Lambda+\setlength{\fboxsep}{0.4mm}\fbox{\scriptsize 2}\ }$ becomes null, or equivalently we should identify $\ket{\Lambda+\setlength{\fboxsep}{0.4mm}\fbox{\scriptsize 1}\ }$ with $\ket{\Lambda+\setlength{\fboxsep}{0.4mm}\fbox{\scriptsize 2}\ }$.}. 
Depending on which pit we fill, we have apparently two different plane partitions, but they should be identical after we apply the periodicity rule.
It may sound that the situation is the same as the one-pit case. The difference is that the existence of two-pits implies that the two diagrams obtained by the translation rule should make sense as the plane partition, which gives an additional restriction on the plane partition with one-pit.


We explain an explicit construction of $p=1,q=2,n=1$ degenerate plane partition which is simple and will be important later.  We have two pits at $(2,3,1)$ and $(1,1,2)$.
The following figure shows the two plane partitions which should be identified.

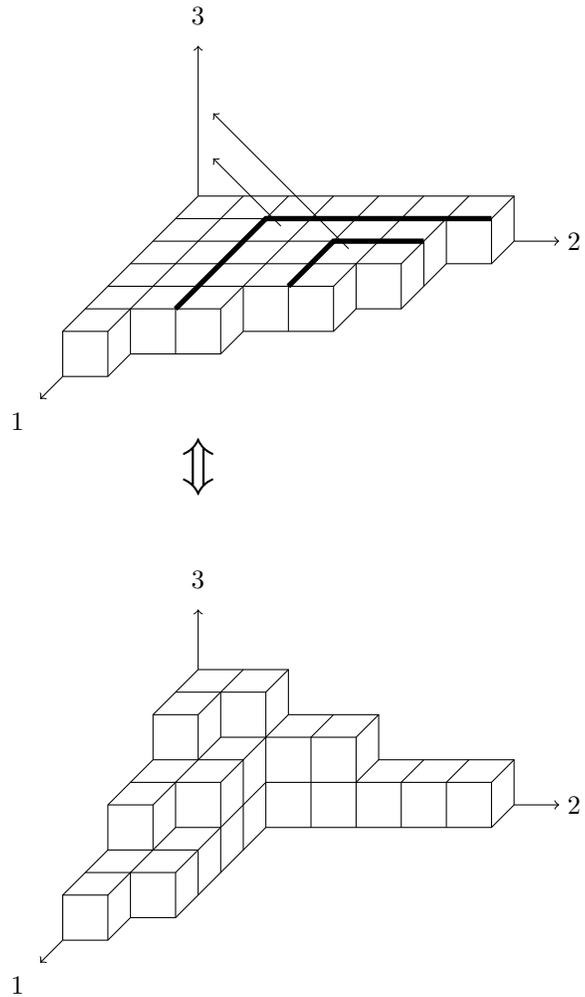
\begin{center}
	\begin{tikzpicture}
	\footnotesize
	
	\draw (-1.8,-1.8)--(-1.2,-1.8)--(-0.9,-1.5)--(0.3,-1.5)--(0.6,-1.2)--(1.8,-1.2)--(2.1,-0.9)--(2.7,-0.9)--(3.3,-0.3)--(3.9,-0.3)--(4.2,0);
	\draw (-1.8,-1.2)--(-1.2,-1.2)--(-0.9,-0.9)--(0.3,-0.9)--(0.6,-0.6)--(1.8,-0.6)--(2.1,-0.3)--(2.7,-0.3)--(3.3,0.3)--(3.9,0.3)--(4.2,0.6);
	\draw (-1.8,-1.8)--(-1.8,-1.2)--(0,0.6);
	\draw (-1.2,-1.8)--(-1.2,-1.2);
	\draw (-0.9,-1.5)--(-0.9,-0.9)--(0.6,0.6);
	\draw (-0.3,-1.5)--(-0.3,-0.9)--(1.2,0.6);
	\draw (0.3,-1.5)--(0.3,-0.9);
	\draw (0.6,-1.2)--(0.6,-0.6)--(1.8,0.6);
	\draw (1.2,-1.2)--(1.2,-0.6)--(2.4,0.6);
	\draw (1.8,-1.2)--(1.8,-0.6);
	\draw (2.1,-0.9)--(2.1,-0.3)--(3,0.6);
	\draw (2.7,-0.9)--(2.7,-0.3);
	\draw (3,-0.6)--(3,0);
	\draw (3.3,-0.3)--(3.3,0.3)--(3.6,0.6);
	\draw (3.9,-0.3)--(3.9,0.3);
	\draw (4.2,0)--(4.2,0.6)--(0,0.6);
	
	\draw (-1.5,-0.9)--(-0.9,-0.9);
	\draw (-1.2,-0.6)--(0.6,-0.6);
	\draw (-0.9,-0.3)--(2.1,-0.3);
	\draw (-0.6,0)--(3,0);
	\draw (-0.3,0.3)--(3.3,0.3);
	
	\draw[line width=2pt] (-0.3,-0.9)--(0.9,0.3)--(3.9,0.3);
	\draw[line width=2pt] (1.2,-0.6)--(1.8,0)--(3,0);
	
	\draw[->] (1.1,0.2)--(0.2,1.1);
	\draw[->] (2,-0.1)--(0.2,1.7);
	
	\draw[->] (0,0.6)--(0,2.6);
	\draw[->] (-1.8,-1.8)--(-2.1,-2.1);
	\draw[->] (4.2,0)--(4.8,0);
	\node (h2) at (-2.4,-2.4) {1};
	\node (h3) at (5,0) {2};
	\node (h1) at (0,3) {3};
	

\begin{scope}[yshift=-7.5cm]
	\draw (-1.8,-1.8)--(-1.2,-1.8)--(-0.9,-1.5)--(-0.3,-1.5)--(0.9,-0.3)--(3.9,-0.3)--(4.2,0);
	\draw (-1.8,-1.2)--(-1.2,-1.2)--(-0.9,-0.9)--(-0.3,-0.9)--(0.9,0.3)--(3.9,0.3)--(4.2,0.6);
	
	\draw (-1.2,-0.6)--(-0.6,-0.6)--(-0.3,-0.3)--(0.3,-0.3);
	\draw (2.1,0.3)--(2.4,0.6);
	\draw (-1.2,0)--(-0.6,0)--(-0.3,0.3)--(0.3,0.3)--(0.9,0.9)--(2.1,0.9)--(2.4,1.2);
	
	\draw (-0.6,0.6)--(0,0.6)--(0.3,0.9)--(0.9,0.9)--(1.2,1.2);
	\draw (-0.6,1.2)--(0,1.2)--(0.3,1.5)--(0.9,1.5)--(1.2,1.8);
	
	\draw (0,1.8)--(-0.6,1.2)--(-0.6,0.6)--(-1.2,0)--(-1.2,-0.6)--(-1.8,-1.2)--(-1.8,-1.8);
	\draw (0,1.8)--(1.2,1.8)--(1.2,1.2)--(2.4,1.2)--(2.4,0.6)--(4.2,0.6)--(4.2,0);
	
	\draw (0.6,1.8)--(0.3,1.5)--(-0.3,1.5);
	\draw (0.3,1.5)--(0.3,0.9);
	\draw (0.9,1.5)--(0.9,0.9);
	\draw (0,1.2)--(0,0.6);
	\draw (-0.3,0.3)--(0,0.6)--(0.6,0.6);
	\draw (-0.9,0.3)--(-0.3,0.3)--(-0.3,-0.3);
	\draw (-0.6,0)--(-0.6,-0.6)--(-0.9,-0.9)--(-1.5,-0.9);
	
	\draw (-1.2,-1.8)--(-1.2,-1.2);
	\draw (-0.9,-1.5)--(-0.9,-0.9);
	\draw (-0.3,-1.5)--(-0.3,-0.9);
	\draw (0,-1.2)--(0,-0.6)--(-0.6,-0.6);
	\draw (0.3,-0.9)--(0.3,0.3);
	\draw (0.6,-0.6)--(0.6,0.6);
	\draw (0.9,-0.3)--(0.9,0.9);
	\draw (1.5,-0.3)--(1.5,0.9)--(1.8,1.2);
	\draw (2.1,-0.3)--(2.1,0.9);
	\draw (2.7,-0.3)--(2.7,0.3)--(3,0.6);
	\draw (3.3,-0.3)--(3.3,0.3)--(3.6,0.6);
	\draw (3.9,-0.3)--(3.9,0.3);
	
	\draw[->] (0,1.8)--(0,2.6);
	\draw[->] (-1.8,-1.8)--(-2.1,-2.1);
	\draw[->] (4.2,0)--(4.8,0);
	\node (h2) at (-2.4,-2.4) {1};
	\node (h3) at (5,0) {2};
	\node (h1) at (0,3) {3};
	
	\node (a) at (0,4.5) {\LARGE$\Updownarrow$};
\end{scope}

	\end{tikzpicture}
\end{center}
The first figure is the case when we fill the pit at $(2,3,1)$. We cannot fill the other pit at $(1,1,2)$, and we obtain a plane partition whose height is one.  The periodicity constraint implies that one can slice the diagram at $(k+1,2k+1,1)$ $k=1,2,\cdots$ in the shape of hook and pile the $(k+1)$-th piece  on the $k$-th piece, which gives a diagram where we identify $(2,3,1)$ as the pit. In this simple example $n=1$, one may take the first diagram as the arbitrary partition.  The second diagram, obtained in this way, belongs to a restricted set of the plane partition with the pit at $(2,3,1)$. For the general $n$, we need restrictions on both diagrams as we will see in the minimal model of $W_n$ algebra.

\section{CFT data associated with asymptotic Young diagrams}
\label{subsubsec:degeneratemodule}
When the plane partition has infinite legs, the Young diagrams which appear in the asymptotes label the representation of the affine Yangian.  In the following, we focus on the case that the number of the directions where non-trivial asymptotes are imposed is at most two. This situation is enough to analyze minimal models later.

\subsection{Conformal dimension and $U(1)$-charge}
We can evaluate the conformal dimension $h$ and $U(1)$ charge $j'$ for the plane partition with non-trivial asymptotic condition by (\ref{eq:psieigenvalue}) and  (\ref{eq:zeromode}). 
We note that the zero modes of $U(1)$ current and Virasoro algebra are identified as follows \cite{Prochazka:2015deb}: 
\begin{equation}
\label{eq:zeromode}
\begin{split}
&J_0\quad\longleftrightarrow\quad\psi_1,\\
&L_0\quad\longleftrightarrow\quad\frac{1}{2}\psi_2,
\end{split}
\end{equation}
where the normalization of $U(1)$ current is 
\begin{equation}
[J_n,J_m]=\psi_0n\delta_{n+m,0}.
\end{equation}
Decoupling $U(1)$ factor, the zero mode of Virasoro algebra is given by 
\begin{equation}
\label{eq:decoupledim}
L_0^{\rm decouple}=\frac{1}{2}\psi_2-\frac{1}{2\psi_0}\psi_1^2.
\end{equation}

One can compute the conformal dimension and $U(1)$ charge for the configuration with non-trivial asymptotes by multiplying an infinite number of $\varphi(u-h_{\fbox{}})$ as in (\ref{eq:psieigenvalue}).
The general formula was given in \cite{Prochazka:2017qum}.  If  the asymptotic Young diagram $(\mu_1,\mu_2,\cdots\mu_l)$ is imposed in $x_2$ direction,  the conformal dimension and $U(1)$ charge are\footnote{In this formula, we use  $\psi_1$ as $U(1)$ zero mode. We will change the normalization later.}
\begin{equation}
\label{eq:formula}
\begin{split}
j'(M_{\mu}^2)&=-\frac{1}{h_2}\sum_j\mu_j,\\
h(M_{\mu}^2)&=-\frac{\lambda_2}{2\lambda_3}\sum_j{\mu_j^2}-\frac{\lambda_2}{2\lambda_1}\sum_j(2j-1)\mu_j+\frac{\lambda_2}{2}\sum_j\mu_j\\
&=-\frac{\lambda_2}{2\lambda_3}\sum_j{\mu_j^2}-\frac{\lambda_2}{2\lambda_1}\sum_j(\mu^T)_j^2+\frac{\lambda_2}{2}\sum_j\mu_j,
\end{split}
\end{equation}
where $\mu^T$ is a transposition of $\mu$ and we take $x_1$ direction as the one associated with $l$.
The other cases can be  understood just by permuting $\lambda_1, \lambda_2$, and $\lambda_3$.  In the case where non-trivial asymptotic conditions are imposed in more than one direction, $U(1)$ charge can be obtained just by summing each factor. For conformal dimension, we need to subtract the number of overlapping boxes in addition to summing each factor. That can be understood intuitively by considering that Virasoro zero mode counts the number of boxes. For later use, we explicitly write the expression for the configuration with asymptotic Young diagram $(\mu_1,\mu_2,\cdots\mu_l)$ in $x_1$ direction and $(\nu_1,\nu_2,\cdots\nu_m)$ in  $x_2$ direction,
\ba
\begin{split}
&j'=j'(M_{\mu}^1)+j'(M_{\nu}^2),\\
&h=h(M_{\mu}^1)+h(M_{\nu}^2)-\#(\mu\cap\nu).
\end{split}
\ea
Here, $\#(\mu\cap\nu)$ represents the number of the overlapping boxes appearing when two asymptotic condition, $\mu$ and $\nu$, are imposed.


\subsection{Analytic continuation to negative weight}

For the application to WoW, we will have to include the negative weight as we have seen even in the free case where the balance of the ghost charges $h$ can be either positive or negative. We need to generalize the above description to the negative weight by the analytic continuation. 
We consider the case that there is a pit at $(L+1,M+1,1)$ and asymptotic Young diagram $(\mu_1,\cdots,\mu_M)$ in $x_1$ direction and $(\nu_1,\cdots,\nu_L)$ in $x_2$ direction. 
 We suppose $\mu_M<0$ and $\mu_M<\nu_L$. To interpret these Young diagrams as ordinal ones which do not contain negative rows, we shift the origin by $\mu_M$ to $x_3$ direction. Then the asymptotic Young diagrams are given by $(\mu_1-\mu_M,\cdots, \mu_{M-1}-\mu_M,0)$ and $(\nu_1-\mu_M,\cdots, \nu_{L}-\mu_M)$. Recalling that the number given by (\ref{eq:assignedvalue}) is assigned to each box, we  see that the effect by shifting the origin appears in the representation theory as $q=\mu_Mh_3$. Summarizing, we  decompose the weight   $(\mu_1,\cdots,\mu_M)$ into  $(\mu_1,\cdots,\mu_1)$ and  $(\mu_1-\mu_M,\cdots\mu_{M-1}-\mu_M,0)$, the first factor corresponding to $U(1)$ factor which plays a role of shifting the origin and the second factor corresponding to the asymptotic Young diagram. One can intuitively see that just as inserting rows with an infinite number of  anti-boxes.
We derive the formula for $h$ and $j'$ in the presence of negative weight. 
Due to the shift of the origin, the eigenvalue of $\psi(u)$ changes as follows: 
\ba
\psi(u)\to\psi(u-\mu_Lh_3)=1+\frac{\psi_0\sigma_3}{u}+\frac{(\psi_1+\psi_0\mu_Lh_3)\sigma_3}{u^2}+\frac{(\psi_2+2\psi_1\mu_Lh_3-\frac{\mu_L^2\lambda_1\lambda_2}{\lambda_3})\sigma_3}{u^3}+\cdots.
\ea
Then the shift of $h$ and $j'$ is 
\ba
\begin{split}
	&j'\to j'+\psi_0\mu_Lh_3,\\
	&h\to h+j'\mu_Lh_3-\frac{\mu_L^2\lambda_1\lambda_2}{2\lambda_3}.
\end{split}
\ea
As a result, $U(1)$ charge and the conformal dimension are
\ba
\label{eq:formula2}
\begin{split}
	j'&=j'(M_{\tilde{\mu}}^1)+j'(M_{\tilde{\nu}}^2)+\psi_0\mu_Lh_3\\&=j'(M_{\mu}^1)+j'(M_{\nu}^2),\\
	h&=h(M_{\tilde{\mu}}^1)+h(M_{\tilde{\nu}}^2)+(j'(M_{\tilde{\mu}}^1)+j'(M_{\tilde{\nu}}^2))\mu_Lh_3-\frac{\mu_L^2\lambda_1\lambda_2}{2\lambda_3}-\#(\tilde{\mu}\cap\tilde{\nu})\\
	&=h(M_{\mu}^1)+h(M_{\nu}^2)-\#(\tilde{\mu}\cap\tilde{\nu})-LM\mu_L,
\end{split}
\ea
where $\tilde{\mu}$ and $\tilde{\nu}$ represent the Young diagrams $(\mu_1-\mu_M,\cdots, \mu_{M-1}-\mu_M,0)$ and $(\nu_1-\mu_M,\cdots, \nu_{L}-\mu_M)$ respectively. 
Note that we use the relation $\frac{L}{\lambda_1}+\frac{M}{\lambda_2}=1$ in the above. This formula shows that we don't need to change the formula for $j'$, but need to add the factor "$-LM\mu_L$" to that for $h$. One may  intuitively use the original formula (\ref{eq:formula}) by considering negative rows as anti-rows   as follows. Let's consider the simplest case  where $\mu_1=\mu_2=\cdots=\mu_M<0$ and $\nu_1=\nu_2=\cdots=\nu_L=0$. One can interpret the configuration as the one where anti-boxes are inserted at $(x_1,x_2,x_3)$ satisfying $x_1\geq L+1$, $1\leq x_2\leq M$ and $\mu_L+1\leq x_3\leq0$. On the other hand, the anti-rows are naturally defined as rows with anti-boxes at   $(x_1,x_2,x_3)$ satisfying $x_1\geq 1$, $1\leq x_2\leq M$ and $\mu_L+1\leq x_3\leq0$. The difference gives the factor "$-LM\mu_L$".

\section{Description of W-algebra minimal models}
\label{sec:w-minimal}
$W_N$ algebra together with $U(1)$ factor corresponds to $Y_{0,0,N}[\Psi]$. 
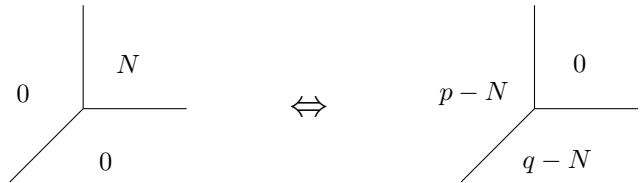
\begin{center}
\begin{tikzpicture}
\footnotesize
\node (D5) at (1.5,0) {};
\node (NS5) at (0,1.5){};
\node (dionic5) at (-1.1,-1.1) { };
\draw (0,0)--(D5);
\draw (0,0)--(NS5);
\draw (0,0)--(dionic5);
\node (D3L) at (-0.8,0.2) {0};
\node (D3M) at (0.3,-0.7) {0};
\node (D3N) at (0.6,0.6) {$N$};
\large
\node (times) at (3,0) {$\Leftrightarrow$};
\scriptsize
\begin{scope}[xshift=6cm,yshift=0cm]
\footnotesize
\node (D5) at (1.5,0) {};
\node (NS5) at (0,1.5){};
\node (dionic5) at (-1.1,-1.1) { };
\draw (0,0)--(D5);
\draw (0,0)--(NS5);
\draw (0,0)--(dionic5);
\node (D3L) at (-0.8,0.2) {$p-N$};
\node (D3M) at (0.3,-0.7) {$q-N$};
\node (D3N) at (0.6,0.6) {0};

\end{scope}

\end{tikzpicture}
\end{center}
In the parametrization of Proch\'azka, it corresponds to $\lambda_3=N$
which satisfies
\ba
\frac{0}{\lambda_1}+\frac{0}{\lambda_2}+\frac{N}{\lambda_3}=1\,,
\ea
for arbitrary choice of $\lambda_1,\lambda_2$ as long as they satisfy (\ref{eq:sumhlam}).
One may parameterize them as $\lambda_1=N(\beta-1)$ and $\lambda_2=\frac{N(1-\beta)}{\beta}$ with $\beta = \Psi^{-1}$.
The central charge of the algebra (\ref{eq:Winftyparameter}) becomes,
\ba\label{eq:Wcenter}
c=(N-1)\left(1-Q^2 N(N+1)\right)+1\,,\quad Q=\sqrt{\beta}-1/\sqrt{\beta},
\ea
which coincides with the known formula up to the second term $1$ which comes the extra $U(1)$. In the plane partition, we have a pit at $(1,1,N+1)$.  The corresponding plane partition consists of $N$-tupple Young diagrams $Y_1,\cdots, Y_N$ with the inclusion relatiosn $Y_i\succ Y_{i+1}$ ($i=1,\cdots, N-1$).  The partition function is
\begin{eqnarray}
\chi(q)=\sum_{n=1}^\infty \sum_{i=0}^{N-1}\frac{1}{(1-q^{n+i})}\,.
\end{eqnarray}

An affine Yangian description of the minimal model was studied in \cite{fukuda2015sh}. Here we give an alternative picture by PP. In addition to the first pit, we introduce the second pit to describe the minimal model. For that purpose we set $\beta=p/q$ ($p,q$ are coprime integers greater than $N$). With this choice, the central charge (\ref{eq:Wcenter}) is identical to those of minimal models.
At the same time, the tuned parameters satisfy
\ba
\frac{p-N}{\lambda_1}+\frac{q-N}{\lambda_2}+\frac{0}{\lambda_3}=1\,,
\ea
which implies that there is the second pit at $(p-N+1, q-N+1,1)$\footnote{We note that not all double truncations give minimal models because $p$ and $q$ must be coprime.}.

The periodicity argument in the previous section implies that it is necessary to identify the boxes at $(x,y,z+N)\sim (x+p-N, y+q-N,z)$.
We divide the set of Young diagrams $Y_i$ into a set of hooks $Y_i^{(k)}$
($k=0,1,2,\cdots$) which are specified by $D_k\setminus D_{k+1}$ with 
$D_k:=\left\{x,y|x>(p-N)k, y>(q-N)k\right\}$.
The dual plane partition after the translation becomes,
$$
Y_1^{(0)}\succeq Y_{2}^{(0)}\succeq \cdots Y_N^{(0)}\succeq Y_1^{(1)}\succeq \cdots \succeq Y_N^{(1)}\succeq Y_1^{(2)}\succeq \cdots.
$$
Such a plane partition is sensible only if $Y_N^{(k)}\succeq Y_1^{(k+1)}$, ($k\geq 0$).  They give extra constraints to the original $N$-tupple Young diagrams.

We note that the primary fields (or the irreducible representations) are parameterized by two set of positive integers $\vec n=(n_1,\cdots, n_{N-1})$ and $\vec n'=(n'_1,\cdots, n'_{N-1})$, $n_i, n'_i\geq 1$ with $\sum_{i=1}^{N-1} n_i<q$ and $\sum_{i=1}^{N-1} n'_i<p$.  The conformal dimension of primary fields is written as,
\ba\label{eq:cdim}
\Delta(\vec n, \vec n')=\frac{12(\sum_{i=1}^{N-1}(pn_i-qn'_i)\vec \omega_i)^2-N(N^2-1)(p-q)^2}{24pq}
\ea 
where $\vec \omega_i$ is the fundamental weight of $su(N)$.
One can describe the module specified by $\vec n, \vec n'$ by including the nonvanishing asymptotic Young diagrams $\mu$ (resp. $\mu'$) in $x_1$ (resp. $x_2$) directions.\footnote{There is no room to include the asymptotic Young diagram in $x_3$ direction since it can not appear in the picture where there is a pit at $(1,1, N+1)$.}
They are defined by,
\ba
\mu&=& (\mu_1,\cdots, \mu_{N-1}),\qquad \mu_j-\mu_{j+1}= n_j-1\\
\mu'&=& (\mu'_1,\cdots, \mu'_{N-1}),\qquad \mu'_j-\mu'_{j+1}={n}'_j-1\,.
\ea
The mutual locations of the Young diagrams can be written in the following form, 
\begin{figure}[H]
	\begin{center}
		\begin{tikzpicture}
		\draw (0,0)--(3.6,0)--(3.6,-0.4)--(2.8,-0.4)--(2.8,-1.2)--(1.6,-1.2)--(1.6,-2.4)--(0.8,-2.4)--(0.8,-3.2)--(0,-3.2)--(0,0);
		\draw[dotted] (0,-3.2)--(0,-4.4);
		\draw[dotted] (3.6,0)--(5.2,0);
		\begin{scope}[xshift=1.2cm,yshift=-0.8cm]
		\draw (1.6,0)--(3.6,0)--(3.6,-0.8)--(2.4,-0.8)--(2.4,-1.6)--(1.6,-1.6)--(1.6,-2.4)--(0.8,-2.4)--(0.8,-3.2)--(0,-3.2)--(0,-1.6);
		\draw[dotted] (0,-3.2)--(0,-4.4);
		\draw[dotted] (3.6,0)--(5.2,0);
		\begin{scope}[xshift=1.2cm,yshift=-0.4cm]
		\draw (2.4,0)--(3.6,0)--(3.6,-0.8)--(2.4,-0.8)--(2.4,-1.6)--(1.6,-1.6)--(1.6,-2.4)--(0.8,-2.4)--(0.8,-3.2)--(0.4,-3.2)--(0.4,-3.6)--(0,-3.6)--(0,-2);
		\draw[dotted] (0,-3.6)--(0,-4);
		\draw[dotted] (3.6,0)--(4,0);
		\begin{scope}[xshift=2.4cm,yshift=-1.6cm]
		\draw (1.2,0)--(3.6,0)--(3.6,-0.8)--(2.4,-0.8)--(2.4,-1.6)--(1.6,-1.6)--(1.6,-2.4)--(0.8,-2.4)--(0.8,-3.2)--(0,-3.2)--(0,-1.2);
		\draw[dotted] (0,-3.2)--(0,-4.4);
		\draw[dotted] (3.6,0)--(4.4,0);
		\begin{scope}[xshift=1.2cm,yshift=-1.2cm]
		\draw (1.2,0)--(2.8,0)--(2.8,-0.8)--(2,-0.8)--(2,-1.2)--(1.6,-1.2)--(1.6,-2)--(0.8,-2)--(0.8,-2.8)--(0,-2.8)--(0,-1.2);
		\draw[dotted] (0,-2.8)--(0,-3.2);
		\draw[dotted] (2.8,0)--(3.2,0);
		\end{scope}
		\end{scope}
		\end{scope}
		\end{scope}
		\node (dot) at (5.3,-3.3) {$\ddots$};
		\draw[->] (-0.3,0.3)--(-0.3,-0.2);
		\draw[->] (-0.3,0.3)--(0.2,0.3);
		\node (1) at (-0.3,-0.4) {1};
		\node (2) at (0.4,0.3) {2}; 
		
		\node (y1) at (1,-1.1) {$Y_N$};
		\node (y2) at (2.5,-2) {$Y_{N-1}$};
		\node (y3) at (3.5,-3) {$Y_{N-2}$};
		\node (y4) at (5.7,-4.5) {$Y_{2}$};
		\node (y5) at (7,-5.3) {$Y_1$};
		
		\footnotesize
		\draw[<->] (0,-4.4)--(1.2,-4.4);
		\node (n1) at (0.7,-4.8) {$\tilde{n}_{N-1}$};
		
		\draw[<->] (1.2,-5.2)--(2.4,-5.2);
		\node (n2) at (1.9,-5.6) {$\tilde{n}_{N-2}$};
		
		\draw[<->] (4.8,-7.2)--(6,-7.2);
		\node (n3) at (5.5,-7.6) {$\tilde{n}_1$};
		
		\node (dot) at (3.5,-6.3) {$\ddots$};
		
		\draw[<->] (5.2,0)--(5.2,-0.8);
		\node (n1t) at (5.8,-0.4) {$\tilde{n}_{N-1}'$};
		
		\draw[<->] (6.4,-0.8)--(6.4,-1.2);
		\node (n2t) at (7,-1) {$\tilde{n}_{N-2}'$};
		
		\draw[<->] (9.2,-2.8)--(9.2,-4);
		\node (n3t) at (9.6,-3.4) {$\tilde{n}_1'$};
		
		\node (dot) at (7.9,-1.8) {$\ddots$};
		
		\end{tikzpicture}
	\end{center}
	\caption{The Young diagrams associated with minimal model of $W_N$ (and $U(1)$) can be interpreted as plane partition by stacking them as shown in the above figure.\label{fig:minimalW}}
\end{figure}
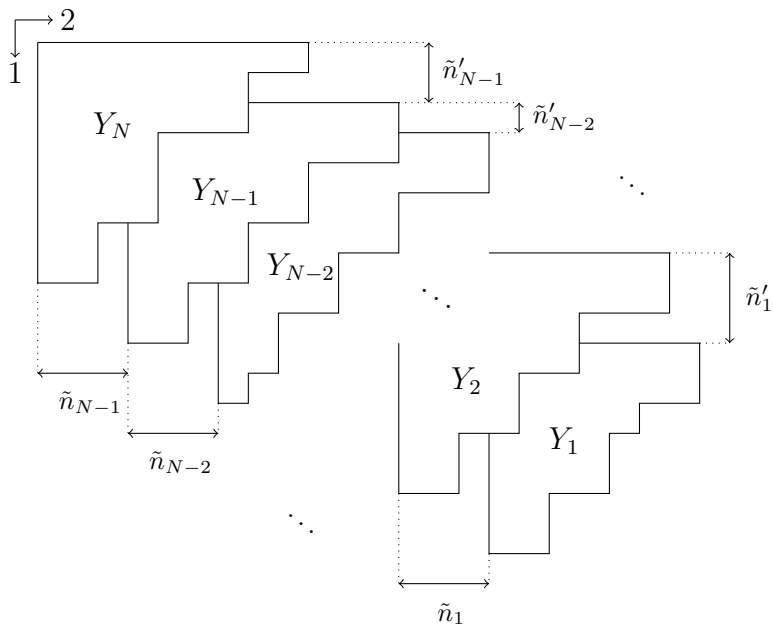

\noindent
The conditions that such configuration makes sense in the sense of plane partition (in both pictures) coincide with the $N$-Burge conditions \cite{Burge:1993,Belavin:2015ria,Alkalaev:2014sma}, which characterize the degenerate module of the minimal models,
\begin{eqnarray}
Y_{i,R}-Y_{i+1,R+n_i-1}\geq -(n'_i-1)\,,
\end{eqnarray}
with $i=1,\cdots, N$. We identify $Y_{N+1}=Y_1$, $n_N=p-\sum_{i=1}^{N-1}n_i$, $n'_N=q-\sum_{i=1}^{N-1}n'_i$.
The condition for $i=N$ is given by the equivalence of two plane partitions obtained by the translation.  
One can also check that the conformal dimension derived from affine Yangian agrees with (\ref{eq:cdim}). See Appendix \ref{app:Wdim} for detail.

\section{$\mathcal{N}=2$ super Virasoro algebra as WoW}
As claimed in \cite{Gaiotto:2017euk,Prochazka:2017qum}, the following diagram of WoW realizes $\mathcal{N}=2$ super Virasoro algebra $\otimes$ $U(1)$ current.
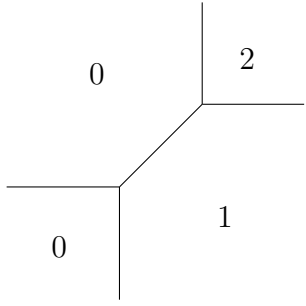
\begin{center}
\begin{tikzpicture}
\node (D5) at (1.5,0) { };
\node (NS5) at (0,1.5) { };
\node (dionic5) at (-1.1,-1.1) { };
\draw (0,0)--(D5);
\draw (0,0)--(NS5);
\draw (0,0)--(-1.1,-1.1);
\draw (-2.6,-1.1)--(-1.1,-1.1);
\draw (-1.1,-2.6)--(-1.1,-1.1);
\node (D3L) at (-1.4,0.4) {0};
\node (D3M) at (0.3,-1.5) {1};
\node (D3N) at (0.6,0.6) {2};
\node (period) at (-1.9,-1.9) {0};
\end{tikzpicture}
\end{center}
In this section, we describe it by the affine Yangian with the emphasis on the shared asymptotic Young diagram, especially the careful treatment of negative weights.

\subsection{WoW set-up}
\label{subsec:interpretation}
The diagram implies that the system consists of $Y_{1,2,0}$ and $Y_{0,0,1}$ with the modules for both of them. 
As explained in subsection \ref{subsec:Y-algebra}, they arise from line operators on the intermediate segment and are characterized by $U(1)$ weight $\mu$, which takes all integer. From the viewpoint of the affine Yangian, 
we can  understand them as plane partitions with a shared asymptotic Young diagram ($\mu$) in the gluing direction. In the following, we refer to them as "intermediate Young diagrams". 

The parameters of two Yangians are,
\begin{equation}
\label{eq:lambda1}
\lambda_1^{(1)}=-\frac{n}{n+2},\quad\lambda_2^{(1)}=\frac{n}{n+1},\quad\lambda_3^{(1)}=n,
\end{equation}
\begin{equation}
\label{eq:lambda2}
\lambda_1^{(2)}=-\frac{1}{n+2},\quad\lambda_2^{(2)}=\frac{1}{n+1},\quad\lambda_3^{(2)}=1.
\end{equation}
The subscripts $(1)$ and $(2)$ at upper right are used to distinguish two Yangians.
 In this parametrization, the gluing direction is the second one. $\Psi$ and $n$ are related by $\Psi=n+2$.
The parameter $n$ can be arbitrary for the description of the general $\mathcal{N}=2$ superconformal algebra. When $n$ becomes positive integers, we have a second pit in $Y^{(1)}$ at $(1,1,n+1)$.  In the next section, we will argue that the system describes the minimal models as in the $W_n$-algebra.

\begin{figure}[H]
\begin{center}
\begin{tikzpicture}
\footnotesize
\node (3) at (4.5,0) {2};
\node (1) at (0,2) {3};
\node (2) at (-2,-2) {1};
\draw[->] (0,0)--(1);
\draw[->] (0,0)--(2);
\draw[->] (0,0)--(3);
\draw (-1.2,-1.2)--(-1.2,-0.6)--(0,0.6);
\draw (0,0.6)--(3.6,0.6);
\draw (-1.2,-0.6)--(2.4,-0.6);
\draw (-1.2,-1.2)--(2.4,-1.2);
\fill[pattern=north west lines] (0,0.6)--(-1.2,-0.6)--(2.2,-0.6)--(3.4,0.6)--cycle;
\fill[pattern=north east lines] (-1.2,-0.6)--(2.2,-0.6)--(2.2,-1.2)--(-1.2,-1.2)--cycle;
\node (mu) at (-1.4,-1.2) {$\mu$};
\node (comment1) at (0.3,-2.8) {(i)\ $\mu>0$};

\node (33) at (13,1) {2};
\node (11) at (10,3) {3};
\node (22) at (7,-2) {1};
\draw[dotted][->] (9,0)--(9,2);
\draw[->] (9,0)--(22);
\draw[dotted][->] (9,0)--(12,0);
\draw (9,0)--(10,1);
\draw[->] (10,1)--(33);
\draw[->] (10,1)--(11);
\node (comment1) at (9.3,-2.8) {(ii)\ $\mu<0$};

\node (pit1) at (0,0.75) {$\times$};
\node (pit2) at (10,1.75) {$\times$};

\end{tikzpicture}
\end{center}
\caption{The left figure shows $\mu>0$ case where asymptotic Young diagram $(\mu)$ is imposed. The right one shows $\mu<0$ case where the origin is shifted to the $x_1$ direction by $\mu$.  The crosses mean pits.
\label{fig:bi-module1}}
\end{figure}
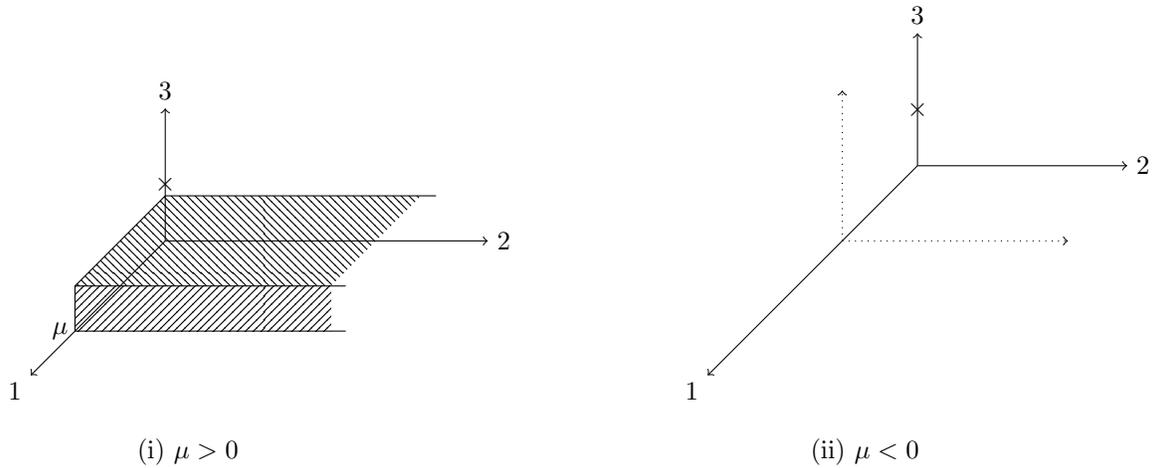

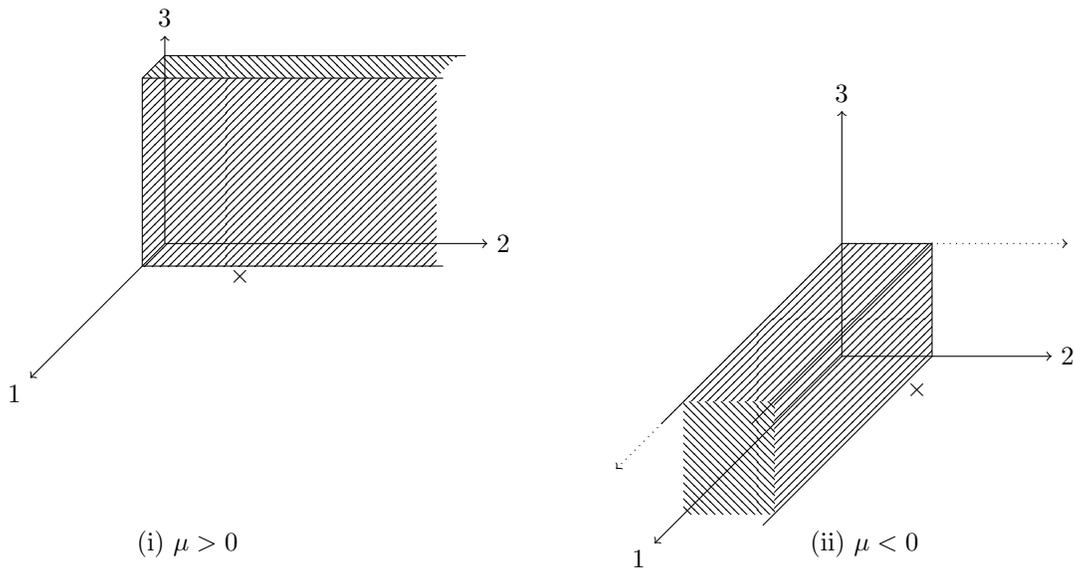
\begin{figure}[H]
\begin{center}
\begin{tikzpicture}
\footnotesize
\node (3) at (4.5,0) {2};
\node (1) at (0,3) {3};
\node (2) at (-2,-2) {1};
\draw[->] (0,0)--(1);
\draw[->] (0,0)--(2);
\draw[->] (0,0)--(3);
\draw (-0.3,-0.3)--(-0.3,2.2)--(0,2.5)--(4,2.5);
\draw (-0.3,2.2)--(3.7,2.2);
\draw (-0.3,-0.3)--(3.7,-0.3);

\node (comment1) at (0.3,-4) {(i)\ $\mu>0$};
\fill[pattern=north west lines] (3.6,2.2)--(-0.3,2.2)--(0,2.5)--(3.9,2.5)--cycle;
\fill[pattern=north east lines] (3.6,2.2)--(-0.3,2.2)--(-0.3,-0.3)--(3.6,-0.3)--cycle;

\node (33) at (12,-1.5) {2};
\node (11) at (9,2) {3};
\node (22) at (6.3,-4.2) {1};
\draw[->] (9,0)--(11);
\draw[dotted][->] (9,0)--(6,-3);
\draw[dotted][->] (9,0)--(12,0);
\draw (7.8,-2.4)--(10.2,0)--(10.2,-1.5)--(7.95,-3.75);
\fill[pattern=north east lines] (8.1,-2.1)--(10.2,0)--(10.2,-1.5)--(8.1,-3.6)--cycle;
\fill[pattern=north east lines] (8.1,-2.1)--(10.2,0)--(9,0)--(6.9,-2.1)--cycle;
\fill[pattern=north west lines] (8.1,-2.1)--(6.9,-2.1)--(6.9,-3.6)--(8.1,-3.6)--cycle;

\draw[->] (9,0)--(9,-1.5)--(33);
\draw[->] (9,-1.5)--(22);
\draw (6.6,-2.4)--(9,0)--(10.2,0);

\node (comment1) at (9.3,-4) {(ii)\ $\mu<0$};

\node (pit1) at (1,-0.43) {$\times$};
\node (pit2) at (10,-1.95) {$\times$};

\end{tikzpicture}
\end{center}
\caption{The left figure shows $\mu>0$ case where asymptotic Young diagram is $(\mu)$. The right one shows $\mu<0$ case. The origin is moved to the $x_3$ direction and asymptotic Young diagram is $(|\mu|,|\mu|)$.\label{fig:bi-module2}}
\end{figure}

We give a graphical representation of $Y_{0,0,1}$ (resp. $Y_{1,2,0}$)  in Figure \ref{fig:bi-module1} (resp. Figure \ref{fig:bi-module2}), where we use the treatment of the negative weight in section \ref{subsubsec:degeneratemodule}.


%
\noindent
We have to mention the different appearance between $\mu>0$ case and $\mu<0$ case. One may feel that the figures for $\mu<0$ look strange because it does not match the picture that two plane partitions share an infinite leg. In the free field case, the sign of the weight changes only the direction in which a shared leg extends and not the shape. That reflects the fact that the sign comes just from that of the supercurrents $G^{\pm}(z)$. From the point, it is better to consider the above figures for $\mu<0$ case as "analytic continuation". Although it does not manifestly keep such a picture, we will see that it gives a correct description.

\subsection{Conformal dimension and $U(1)$ charge of intermediate Young diagrams}
\label{subsec:conformaldimcharge}
In section \ref{subsubsec:degeneratemodule},  we derived the conformal weight and the $U(1)$ factor for the affine Yangian.  
The formula (\ref{eq:formula}) or (\ref{eq:formula2}) gives the conformal dimension ($=$ the eigenvalue of $\frac{1}{2}(\psi_2^{(1)}+\psi_2^{(2)})$) of the intermediate Young diagram with the height $\mu$ as
\begin{equation}
\label{eq:totaldim}
h=\frac{|\mu|(|\mu|+2)}{2}.
\end{equation}
For the $U(1)$ charge, we note that there are two $U(1)$ charges coming from two $Y$ algebra.  We need to specify the linear combination of the two which describes the $U(1)$ current of $\mathcal{N}=2$ superconformal algebra (SCA).  The other $U(1)$ current should commute with the operators in $\mathcal{N}=2$ SCA. We will refer to the former (resp. latter) current as $J$ (resp. $j$).

For the simplicity of the notation, we adjust the parameters $h_i^{(a)}$ by using scaling symmetry of the affine Yangian so that $h_i^{(1)}=h_i^{(2)}$ for $i=1,2,3$. In this convention, we may and will omit the subscript in $h_i$. The relation between $\psi_0^{(1)}$ and $\psi_0^{(2)}$ becomes  $\psi_0^{(1)}=n\psi_0^{(2)}$.

We note that the current $j$ should give a vanishing charge for any $\mu$ since the intermediate Young diagrams  are generated by the supercharges $G^+, G^-$, which leads to
\begin{equation}
\label{eq:smallj}
j=\psi_1^{(1)}-\psi_1^{(2)}.
\end{equation}
 
To fix $J$, we impose the charge of the intermediate Young diagram with the weight $\mu$ to be $\mu$, which implies,
\begin{equation}
\label{eq:u1normalization}
J=-h_2\frac{\quad \frac{\psi_1^{(1)}}{\psi_0^{(1)}}+\frac{\psi_1^{(2)}}{\psi_0^{(2)}}\quad }{\quad \frac{1}{\psi_0^{(1)}}+\frac{1}{\psi_0^{(2)}}\quad }=-h_2\frac{\psi_1^{(1)}+n\psi_1^{(2)}}{n+1}.
\end{equation}
We note that this also fits with the definition of $\mathcal{N}=2$ superconformal algebra where the $U(1)$ current is normalized by,
\begin{equation}
\label{eq:standard}
\begin{split}
&[J_n,J_m]=\frac{c}{3}n\delta_{n+m,0},\\
&[J_n,G^{\pm}_r]=\pm G_{n+r}^{\pm}.
\end{split}
\end{equation}
We follow the standard normalization (\ref{eq:standard}) where the central charge   is given from (\ref{eq:Winftyparameter}) by \footnote{We subtract $1$ to decouple $U(1)$ factor.}
\begin{equation}
\label{eq:cc}
c=\frac{3n}{n+2}.
\end{equation}
One can check that the first equation in (\ref{eq:standard}) is also satisfied.

We recall that the state with conformal dimension (\ref{eq:totaldim}) and $U(1)$ charge $\mu$ can be uniquely determined as 
\ba
\label{eq:supercurrentprimary}
\begin{cases}
\prod_{i=1}^{\mu}G_{-i-1/2}^+\ket{0}\quad({\rm for\  }\mu>0)\\
\prod_{i=1}^{-\mu}G_{-i-1/2}^-\ket{0}\quad({\rm for\  }\mu<0).
\end{cases}
\ea
As pointed out in \cite{Prochazka:2017qum}, it corresponds to the primary field $(\partial^{|\mu|-1}G^{\pm}(\partial^{|\mu|-2}G^{\pm}(\cdots(\partial G^{\pm}G^{\pm})\cdots)))(z)$. This gives further confirmation of our convention of $U(1)$ current.

\section{$\mathcal{N}=2$ unitary minimal models}
As one can see from (\ref{eq:cc}), the central charge is equal to that of $\mathcal{N}=2$ unitary minimal model  when we set $n$ to a positive integer.
It is known that $\mathcal{N}=2$ unitary minimal model has the following NS primary fields parametrized by two integers $l,m$:
\ba
\label{eq:n2primary}
\begin{split}
&h_{l,m}=\frac{l(l+2)-m^2}{4(n+2)},\quad J_{l,m}=\frac{m}{n+2},\\
\ \\
&0\leq l\leq n,\quad -l\leq m\leq l,\quad l-m\equiv0\ ({\rm mod}2).
\end{split}
\ea
The characters $\chi(\tau,z):={\rm Tr}\ q^{L_0-\frac{c}{24}}y^{J_0}$ $(q={\rm e}^{2\pi i\tau}, y={\rm e}^{2\pi iz})$ for these primary fields are also known \cite{Ravanini:1987yg,Matsuo:1986cj}:
\ba
\label{eq:NScharacterformula}
\chi_{l,m}(\tau,z)=\sum_{r\in\mathbb{Z}_{2n}}c_{l,m+2r}^{(n)}(\tau)\Theta_{2m+2r(n+2),2n(n+2)}(\tau,\frac{z}{n+2}).
\ea
See Appendix \ref{app:stringfunc} for the definition of  string function $c_{l,m}^{(n)}(\tau)$ and theta function $\Theta_{m,n}(\tau,z)$.
In this section, we show that the WoW correctly reproduce these charges as well as the character of the minimal models.
We note that the Ramond sector is obtained from the NS sector by the spectral flow and we do not need a separate analysis.


\subsection{$n=1$ case}
We first study the simplest nontrivial example $n=1$ since it is illuminating to understand the fundamental rule. The central charge is one
($
c=1
$), 
and we have three irreducible representations,
(i) $l=m=0$, ($h=J=0$), (ii) $l=m=1$, ($h=\frac16, J=\frac13$), (iii) $l=-m=1$, ($h=\frac16, J=-\frac13$).

The novelty appearing in the minimal model is that we have the second pit at $(1,1,2)$. we have the translational periodicity $(x, y,z+1)\sim (x+1,y+2, z)$.  In the previous section, we introduced an infinite leg of the shape $(\mu)$ (resp. $(|\mu|,|\mu|)$) for $\mu>0$ (resp. $\mu<0$) which describe the action of the supercurrents to the vacuum. These legs have the dual descriptions as in section \ref{subsec:truncation}, which impose further constraints on the plane partition.

For $\mu>0$ case, the leg $(\mu)$ is translated to the height one Young table of the following shape,
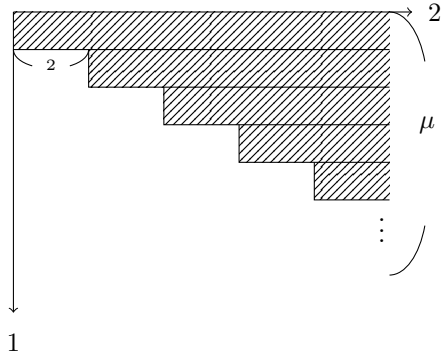
\begin{center}
\begin{tikzpicture}
\footnotesize

\draw[->] (0,0)--(0,-4);
\draw[->] (0,0)--(5.3,0);

\draw (0,-0.5)--(5,-0.5);
\draw (1,-0.5)--(1,-1)--(5,-1);
\draw (2,-1)--(2,-1.5)--(5,-1.5);
\draw (3,-1.5)--(3,-2)--(5,-2);
\draw (4,-2)--(4,-2.5)--(5,-2.5);

\node (3) at (5.6,0) {$2$};
\node (2) at (0,-4.4) {$1$};
\node (dot) at (4.9,-2.8) {$\vdots$};

\fill[pattern=north east lines] (0,0)--(0,-0.5)--(5,-0.5)--(5,0)--cycle;
\fill[pattern=north east lines] (1,-0.5)--(1,-1)--(5,-1)--(5,-0.5)--cycle;
\fill[pattern=north east lines] (2,-1)--(2,-1.5)--(5,-1.5)--(5,-1)--cycle;
\fill[pattern=north east lines] (3,-1.5)--(3,-2)--(5,-2)--(5,-1.5)--cycle;
\fill[pattern=north east lines] (4,-2)--(4,-2.5)--(5,-2.5)--(5,-2)--cycle;

\draw (5,-3.5) arc(270:340:0.5 and 1);
\draw (5,0) arc(90:20:0.5 and 1);
\node (m) at (5.5,-1.5) {$\mu$};

\draw (0,-0.5) arc(180:240:0.5 and 0.25);
\draw (1,-0.5) arc(360:300:0.5 and 0.25);
\node (m) at (0.5,-0.7) {\tiny $2$};

\end{tikzpicture}
\end{center}
Obviously, one can not regard it as a Young diagram with asymptotes unless we fill the vacant boxes located on the left  side of the leg factors.  It implies that the plane partition with the simple leg $(\mu)$ becomes null and we have to add $2+4+\cdots+2(\mu-1)$ boxes in order to realize the non-vanishing states. That can also be seen by explicitly computing the norm. See Appendix \ref{app:a} for detail.

Similarly for $\mu<0$, the infinite leg with the shape $(|\mu|,|\mu|)$ is translated into the following height one diagram,
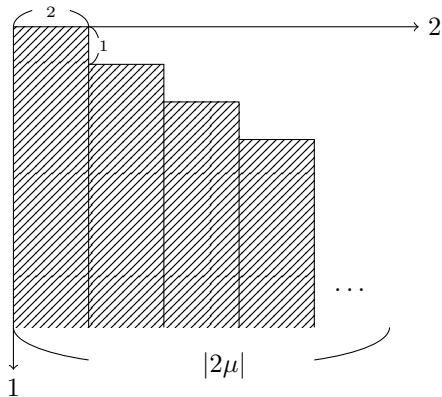
\begin{center}
\begin{tikzpicture}
\footnotesize

\node (3) at (5.6,0) {$2$};
\node (2) at (0,-4.8) {$1$};
\draw[->] (0,0)--(2);
\draw[->] (0,0)--(3);

\draw (1,0)--(1,-4);
\draw (1,-0.5)--(2,-0.5)--(2,-4);
\draw (2,-1)--(3,-1)--(3,-4);
\draw (3,-1.5)--(4,-1.5)--(4,-4);

\node (dot) at (4.5,-3.5) {$\dots$};

\fill[pattern=north east lines] (0,0)--(0,-4)--(1,-4)--(1,0)--cycle;
\fill[pattern=north east lines] (1,-0.5)--(1,-4)--(2,-4)--(2,-0.5)--cycle;
\fill[pattern=north east lines] (2,-1)--(2,-4)--(3,-4)--(3,-1)--cycle;
\fill[pattern=north east lines] (3,-1.5)--(3,-4)--(4,-4)--(4,-1.5)--cycle;

\draw (0,-4) arc(180:240:2 and 0.5);
\draw (5,-4) arc(360:300:2 and 0.5);
\node (m) at (2.8,-4.5) {$|2\mu|$};

\draw (0,0) arc(180:120:0.5 and 0.25);
\draw (1,0) arc(0:60:0.5 and 0.25);
\node (m) at (0.5,0.2) {\tiny $2$};

\draw (1,-0.5) arc(270:330:0.15 and 0.3);
\draw (1,0) arc(90:30:0.15 and 0.3);
\node (m) at (1.2,-0.25) {\tiny $1$};
\end{tikzpicture}
\end{center}
Again the translated legs are not consistent as a Young diagram unless we fill the vacant boxes on the top of the shaded region.

The necessity of adding extra boxes are essential to reproduce the CFT parameters and the characters of the irreducible representations.  The distinction between the different representation is described by the modification of the leg factor in $x_1$ direction.\footnote{One cannot add the infinite leg in $x_3$ direction since there is a pit at $(1,1,2)$. $x_2$ direction is used to describe the intermediate Young diagrams. The freedom exists only in $x_1$ direction.}

\subsubsection{$l=m=0$}
In this case, there is no need to introduce extra asymptotes in $x^1$ direction.
We have to fill extra $\sum_{i=1}^{|\mu |} 2(i-1)=|\mu|(|\mu|-1)$ boxes to describe the nonvanishing state.
Putting this factor and (\ref{eq:totaldim}) together, we  obtain the  character as follows:
\begin{equation}
\begin{split}
\chi(\tau,z)&=\frac{\sum_{\mu=-\infty}^{\infty}q^{\frac{|\mu|(|\mu|+2)}{2}+|\mu|(|\mu|-1)}y^{\mu}}{\eta(\tau)^2}\\
&=\frac{\sum_{\mu=-\infty}^{\infty}q^{\frac{3\mu^2}{2}}y^{\mu}}{\eta(\tau)}\cdot\frac{1}{\eta(\tau)}\\
&=\frac{(\Theta_{0,6}(\tau,\frac{z}{3})+\Theta_{6,6}(\tau,\frac{z}{3}))}{\eta(\tau)}\cdot\frac{1}{\eta(\tau)},
\end{split}
\end{equation}
where we use Dedekind Eta function defined in (\ref{eq:etatheta}).  We note that the character for $Y_{0,0,1}$ does not depend on $\mu$ and is the common second factor (the character for free boson). If we neglect this factor, it exactly coincides the vacuum character of $\mathcal{N}=2$ unitary minimal model in NS sector with $n=1$ \cite{Ravanini:1987yg,Matsuo:1986cj}. 

\subsubsection{$l=1,\ m=-1$}
In addition to intermediate Young diagrams in $x_2$ direction, we introduce an infinite leg with Young diagram $(1)$ in $Y_{1,2,0}$ in $x_1$ direction. 
When $\mu>0$, the infinite legs in the picture with a pit at $(1,1,2)$ appears as
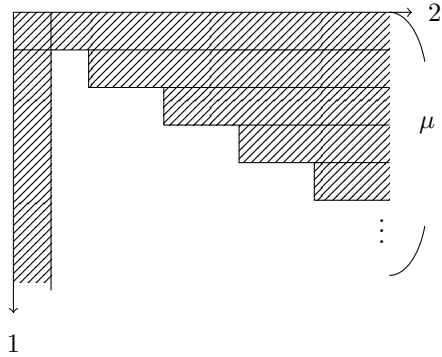
\begin{center} 
\begin{tikzpicture}
\footnotesize

\draw[->] (0,0)--(0,-4);
\draw[->] (0,0)--(5.3,0);

\draw (0,-0.5)--(5,-0.5);
\draw (1,-0.5)--(1,-1)--(5,-1);
\draw (2,-1)--(2,-1.5)--(5,-1.5);
\draw (3,-1.5)--(3,-2)--(5,-2);
\draw (4,-2)--(4,-2.5)--(5,-2.5);
\draw (0.5,0)--(0.5,-3.7);

\node (3) at (5.6,0) {$2$};
\node (2) at (0,-4.4) {$1$};
\node (dot) at (4.9,-2.8) {$\vdots$};

\fill[pattern=north east lines] (0,0)--(0,-0.5)--(5,-0.5)--(5,0)--cycle;
\fill[pattern=north east lines] (1,-0.5)--(1,-1)--(5,-1)--(5,-0.5)--cycle;
\fill[pattern=north east lines] (2,-1)--(2,-1.5)--(5,-1.5)--(5,-1)--cycle;
\fill[pattern=north east lines] (3,-1.5)--(3,-2)--(5,-2)--(5,-1.5)--cycle;
\fill[pattern=north east lines] (4,-2)--(4,-2.5)--(5,-2.5)--(5,-2)--cycle;
\fill[pattern=north east lines] (0.5,-0.5)--(0,-0.5)--(0,-3.6)--(0.5,-3.6)--cycle;

\draw (5,-3.5) arc(270:340:0.5 and 1);
\draw (5,0) arc(90:20:0.5 and 1);
\node (m) at (5.5,-1.5) {$\mu$};

\end{tikzpicture}
\end{center}
As in $l=m=0$ case, we have to fill the open boxes on the right of $\mu$ infinite rows which amounts to $\sum_{i=1}^{\mu-1}(2i-1)=(\mu-1)^2$ boxes.
When $\mu<0$, the state can be identified  as follows:
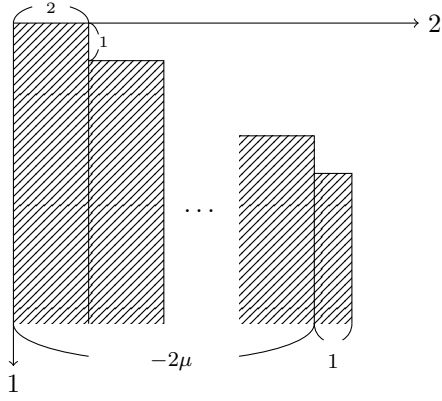
\begin{center}
\begin{tikzpicture}
\footnotesize

\node (3) at (5.6,0) {$2$};
\node (2) at (0,-4.8) {$1$};
\draw[->] (0,0)--(2);
\draw[->] (0,0)--(3);

\draw (1,0)--(1,-4);
\draw (1,-0.5)--(2,-0.5)--(2,-4);
\draw (3,-1.5)--(4,-1.5)--(4,-4);
\draw (4,-2)--(4.5,-2)--(4.5,-4);

\node (dot) at (2.5,-2.5) {$\dots$};

\fill[pattern=north east lines] (0,0)--(0,-4)--(1,-4)--(1,0)--cycle;
\fill[pattern=north east lines] (1,-0.5)--(1,-4)--(2,-4)--(2,-0.5)--cycle;
\fill[pattern=north east lines] (3,-1.5)--(3,-4)--(4,-4)--(4,-1.5)--cycle;
\fill[pattern=north east lines] (4,-2)--(4,-4)--(4.5,-4)--(4.5,-2)--cycle;

\draw (0,-4) arc(180:240:2 and 0.5);
\draw (4,-4) arc(360:300:2 and 0.5);
\node (m) at (2.1,-4.5) {\scriptsize $-2\mu$};

\draw (4,-4) arc(180:240:0.25 and 0.25);
\draw (4.5,-4) arc(360:300:0.25 and 0.25);
\node (m) at (4.25,-4.5) {\scriptsize $1$};

\draw (0,0) arc(180:120:0.5 and 0.25);
\draw (1,0) arc(0:60:0.5 and 0.25);
\node (m) at (0.5,0.2) {\tiny $2$};

\draw (1,-0.5) arc(270:330:0.15 and 0.3);
\draw (1,0) arc(90:30:0.15 and 0.3);
\node (m) at (1.2,-0.25) {\tiny $1$};

\end{tikzpicture}
\end{center}
The number of boxes which we have to fill is $\mu^2$.

$U(1)$ charge is computed from (\ref{eq:formula}) and (\ref{eq:u1normalization}) as
\begin{equation}
J=\mu-\frac{1}{3}.
\end{equation}
We claim that the conformal dimension after filling the boxes is
\ba
\label{eq:coformaldim1}
h=\frac{3\mu^2}{2}-\mu+\frac{1}{3}.
\ea
We demonstrate the formula in $\mu>0$ case. First, the contribution from the infinite leg is $\frac{1}{3}$ by (\ref{eq:formula}). Second, the contribution from the intermediate Young diagram  is $\frac{\mu^2}{2}+\mu$ as in (\ref{eq:totaldim}). Third, we need to subtract the number of overlapping boxes from the sum of each factor, which is the box at the origin in $Y_{1,2,0}$ side in this case. Finally, we have to fill $(\mu-1)^2$ boxes. 
The computation for $\mu<0$ case is similar.  From (\ref{eq:coformaldim1}), we see that the highest weight state is in $\mu=0$ whose conformal dimension and $U(1)$ charge are $h=\frac{1}{3}$ and $J=-\frac{1}{3}$ respectively. 
To explain the discrepancy of the conformal weight in the minimal model, we need to decouple $U(1)$ factor. It is given by (\ref{eq:smallj}), and its Virasoro algebra is as usual given by 
\ba
\label{eq:u1sugawara}
L_0^{U(1)}=\frac{(\psi_1^{(1)}-\psi_1^{(2)})^2}{2(\psi_0^{(1)}+\psi_0^{(2)})}.
\ea
In the case under consideration, 
\ba
L_0^{U(1)}=\frac{1}{4h_1^2\psi_0^{(1)}}=\frac{h_2h_3}{4\psi_0^{(1)}\sigma_3h_1}=-\frac{\lambda_1^{(1)}}{4\lambda_2^{(1)}\lambda_3^{(1)}}=\frac{1}{6}.
\ea
After removing the contribution from $U(1)$ factor, the conformal dimension and $U(1)$ charge of the highest weight state are equal to
\ba
h=\frac{1}{6},\quad J=-\frac{1}{3},
\ea
which agree with the unitary minimal model. 
The character can be computed in the similar way with the vacuum case. 
\begin{equation}
\begin{split}
\chi(\tau,z)
&=\frac{\sum_{\mu=-\infty}^{\infty}q^{\frac{3}{2}(\mu-\frac{1}{3})^2}y^{\mu-\frac{1}{3}}}{\eta(\tau)}\cdot\frac{q^{\frac{1}{6}}}{\eta(\tau)}\\
&=\frac{(\Theta_{-2,6}(\tau,\frac{z}{3})+\Theta_{4,6}(\tau,\frac{z}{3}))}{\eta(\tau)}\cdot\frac{q^{\frac{1}{6}}}{\eta(\tau)}
\end{split}
\end{equation}
which agrees with literature. 

\subsubsection{$l=m=1$}
The asymptotic Young diagram becomes $(1,1)$ on $Y_{1,2,0}$. The number of missing boxes necessary is $(\mu-2)(\mu-1)$ in $\mu>0$ case and $\mu^2-\mu$ in $\mu<0$ case. 
After short computation, we have
\begin{equation}
J=\mu-\frac{2}{3},\qquad L_0^{U(1)}=\frac{2}{3},\qquad
h=\frac{3\mu^2}{2}-2\mu+\frac{4}{3}.
\end{equation}
We see that the highest weight state corresponds to $\mu=1$. The conformal dimension and $U(1)$ charge after decoupling $U(1)$ factor  is  $h=\frac{1}{6}$ and  $J=\frac{1}{3}$, and these results indeed correspond to the primary field in NS sector. The character given below agrees with the literature.
\begin{equation} 
\begin{split}  
\chi(\tau,z)&=\frac{\sum_{\mu=-\infty}^{\infty}q^{\frac{3}{2}(\mu-\frac{2}{3})^2}y^{\mu-\frac{2}{3}}}{\eta(\tau)}\cdot\frac{q^{\frac{2}{3}}}{\eta(\tau)}\\
&=\frac{(\Theta_{2,6}(\tau,\frac{z}{3})+\Theta_{-4,6}(\tau,\frac{z}{3}))}{\eta(\tau)}\cdot\frac{q^{\frac{2}{3}}}{\eta(\tau)}.
\end{split}
\end{equation}


\subsection{General $n$}
We first analyze the conformal weight and $U(1)$ charge of the configuration with asymptotic Young diagram $(\nu_1,\nu_2)$ in $x_1$ direction of $Y_{1,2,0}$.
When the weight of the intermediate Young diagram is zero ($\mu=0$), (\ref{eq:formula}) gives, 
\ba
\label{eq:conformaldimgeneral}
h=\frac{\nu_1^2+\nu_2^2+\nu_1+(2n+3)\nu_2}{2(n+2)},\quad J=-\frac{\nu_1+\nu_2}{n+2}.
\ea
We need to subtract the conformal weight by the $U(1)$ part, (\ref{eq:u1sugawara}),
\ba
L_0^{U(1)}=\frac{(\nu_1+\nu_2)^2}{2(n+2)}
\ea
which gives,
\ba 
h=\frac{-2\nu_1\nu_2+\nu_1+(2n+3)\nu_2}{2(n+2)}.
\ea
The highest weight state lies in $\mu=1$ if $\nu_2\geq1$ and $\mu=0$ if $\nu_2=0$ which may be seen by comparing the contribution to the conformal dimension from the intermediate Young diagram with  the one from the overlapping boxes. To summarize, the conformal dimension and $U(1)$ charge of the highest weight state becomes
\ba
\begin{cases}
&h=\frac{-2\nu_1\nu_2+\nu_1+(2n+3)\nu_2}{2(n+2)}-\frac{1}{2},\quad J=1-\frac{\nu_1+\nu_2}{n+2},\\
&h=\quad\frac{\nu_1}{2(n+2)},\qquad \qquad J=-\frac{\nu_1}{n+2}.
\end{cases}
\ea
which should be compared with (\ref{eq:n2primary}). We find the one-to-one correspondence between $\nu_1, \nu_2$ with $l,m$:
\ba
\label{eq:mapping}
\begin{cases}
&\nu_1=n+1-\frac{l+m}{2},\quad\nu_2=\frac{l-m}{2}+1\qquad({\rm for\ }\nu_2\geq1),\\
&\nu_1=l=-m\hspace{130pt}({\rm for\ }\nu_2=0).
\end{cases}
\ea
It establishes the correspondence between the conformal parameters and the plane partition.

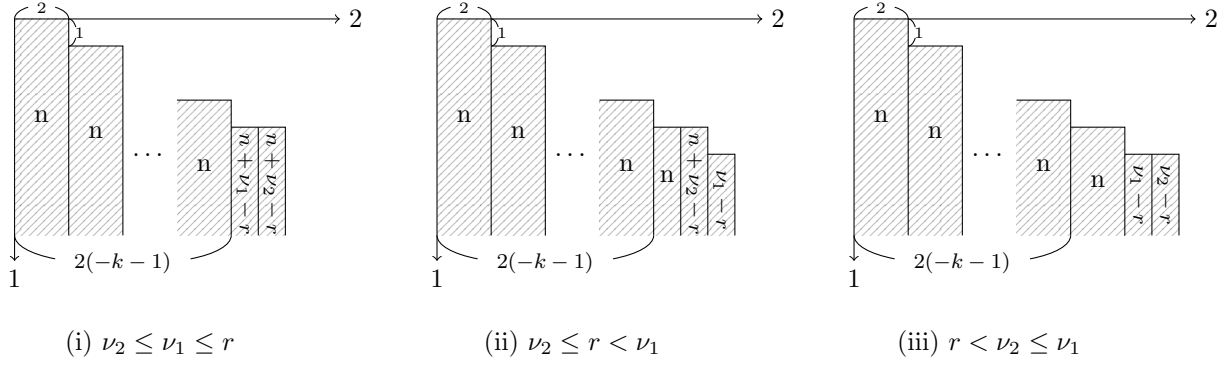
\begin{figure}[H]
\begin{center}
\begin{tikzpicture}[scale=0.72]
\footnotesize

\node (33) at (6.3,0) {$2$};
\node (22) at (0,-4.8) {$1$};
\draw[->] (0,0)--(22);
\draw[->] (0,0)--(33);

\draw (1,0)--(1,-4);
\draw (1,-0.5)--(2,-0.5)--(2,-4);
\draw (3,-1.5)--(4,-1.5)--(4,-4);
\draw (4,-2)--(5,-2)--(5,-4);
\draw (4.5,-2)--(4.5,-4);

\node (dot) at (2.5,-2.5) {$\dots$};

\fill[pattern=north east lines, opacity=0.5] (0,0)--(0,-4)--(1,-4)--(1,0)--cycle;
\fill[pattern=north east lines, opacity=0.5] (1,-0.5)--(1,-4)--(2,-4)--(2,-0.5)--cycle;
\fill[pattern=north east lines, opacity=0.5] (3,-1.5)--(3,-4)--(4,-4)--(4,-1.5)--cycle;
\fill[pattern=north east lines, opacity=0.5] (4,-2)--(4,-4)--(5,-4)--(5,-2)--cycle;

\draw (0,-4) arc(180:240:1.5 and 0.5);
\draw (4,-4) arc(360:300:1.5 and 0.5);
\node (m) at (2,-4.5) {\scriptsize$2(-k-1)$};

\draw (0,0) arc(180:120:0.5 and 0.25);
\draw (1,0) arc(0:60:0.5 and 0.25);
\node (m) at (0.5,0.2) {\tiny $2$};

\draw (1,-0.5) arc(270:330:0.15 and 0.3);
\draw (1,0) arc(90:30:0.15 and 0.3);
\node (m) at (1.2,-0.25) {\tiny $1$};

\node (n1) at (0.5,-1.8) {n};
\node (n2) at (1.5,-2.1) {n};
\node (n3) at (3.5,-2.7) {n};
\node (4) at (4.25,-3.05) {\scriptsize\rotatebox{-90}{$n+\nu_1-r$}};
\node (5) at (4.75,-3.05) {\scriptsize\rotatebox{-90}{$n+\nu_2-r$}};

\node (i) at (2.5,-6) {(i) $\nu_2\leq\nu_1\leq r$};

\begin{scope}[xshift=7.8cm]
\node (33) at (6.3,0) {$2$};
\node (22) at (0,-4.8) {$1$};
\draw[->] (0,0)--(22);
\draw[->] (0,0)--(33);

\draw (1,0)--(1,-4);
\draw (1,-0.5)--(2,-0.5)--(2,-4);
\draw (3,-1.5)--(4,-1.5)--(4,-4);
\draw (4,-2)--(5,-2)--(5,-4);
\draw (4.5,-2)--(4.5,-4);
\draw (5,-2.5)--(5.5,-2.5)--(5.5,-4);

\node (dot) at (2.5,-2.5) {$\dots$};

\fill[pattern=north east lines, opacity=0.5] (0,0)--(0,-4)--(1,-4)--(1,0)--cycle;
\fill[pattern=north east lines, opacity=0.5] (1,-0.5)--(1,-4)--(2,-4)--(2,-0.5)--cycle;
\fill[pattern=north east lines, opacity=0.5] (3,-1.5)--(3,-4)--(4,-4)--(4,-1.5)--cycle;
\fill[pattern=north east lines, opacity=0.5] (4,-2)--(4,-4)--(5,-4)--(5,-2)--cycle;
\fill[pattern=north east lines, opacity=0.5] (5,-2.5)--(5.5,-2.5)--(5.5,-4)--(5,-4)--cycle;

\draw (0,-4) arc(180:240:1.5 and 0.5);
\draw (4,-4) arc(360:300:1.5 and 0.5);
\node (m) at (2,-4.5) {\scriptsize$2(-k-1)$};

\draw (0,0) arc(180:120:0.5 and 0.25);
\draw (1,0) arc(0:60:0.5 and 0.25);
\node (m) at (0.5,0.2) {\tiny $2$};

\draw (1,-0.5) arc(270:330:0.15 and 0.3);
\draw (1,0) arc(90:30:0.15 and 0.3);
\node (m) at (1.2,-0.25) {\tiny $1$};

\node (n1) at (0.5,-1.8) {n};
\node (n2) at (1.5,-2.1) {n};
\node (n3) at (3.5,-2.7) {n};
\node (n4) at (4.25,-2.9) {n};
\node (4) at (5.25,-3.35) {\scriptsize\rotatebox{-90}{$\nu_1-r$}};
\node (5) at (4.75,-3.05) {\scriptsize\rotatebox{-90}{$n+\nu_2-r$}};

\node (i) at (2.5,-6) {(ii) $\nu_2\leq r<\nu_1$};

\end{scope}

\begin{scope}[xshift=15.5cm]
\node (33) at (6.6,0) {$2$};
\node (22) at (0,-4.8) {$1$};
\draw[->] (0,0)--(22);
\draw[->] (0,0)--(33);

\draw (1,0)--(1,-4);
\draw (1,-0.5)--(2,-0.5)--(2,-4);
\draw (3,-1.5)--(4,-1.5)--(4,-4);
\draw (4,-2)--(5,-2)--(5,-4);
\draw (5,-2.5)--(6,-2.5)--(6,-4);
\draw (5.5,-2.5)--(5.5,-4);

\node (dot) at (2.5,-2.5) {$\dots$};

\fill[pattern=north east lines, opacity=0.5] (0,0)--(0,-4)--(1,-4)--(1,0)--cycle;
\fill[pattern=north east lines, opacity=0.5] (1,-0.5)--(1,-4)--(2,-4)--(2,-0.5)--cycle;
\fill[pattern=north east lines, opacity=0.5] (3,-1.5)--(3,-4)--(4,-4)--(4,-1.5)--cycle;
\fill[pattern=north east lines, opacity=0.5] (4,-2)--(4,-4)--(5,-4)--(5,-2)--cycle;
\fill[pattern=north east lines, opacity=0.5] (5,-2.5)--(6,-2.5)--(6,-4)--(5,-4)--cycle;
\draw (0,-4) arc(180:240:1.5 and 0.5);
\draw (4,-4) arc(360:300:1.5 and 0.5);
\node (m) at (2,-4.5) {\scriptsize$2(-k-1)$};

\draw (0,0) arc(180:120:0.5 and 0.25);
\draw (1,0) arc(0:60:0.5 and 0.25);
\node (m) at (0.5,0.2) {\tiny $2$};

\draw (1,-0.5) arc(270:330:0.15 and 0.3);
\draw (1,0) arc(90:30:0.15 and 0.3);
\node (m) at (1.2,-0.25) {\tiny $1$};

\node (n1) at (0.5,-1.8) {n};
\node (n2) at (1.5,-2.1) {n};
\node (n3) at (3.5,-2.7) {n};
\node (n4) at (4.5,-3) {n};
\node (4) at (5.25,-3.25) {\scriptsize\rotatebox{-90}{$\nu_1-r$}};
\node (5) at (5.75,-3.25) {\scriptsize\rotatebox{-90}{$\nu_2-r$}};

\node (i) at (2.5,-6) {(iii) $r<\nu_2\leq\nu_1$};
\end{scope}
\end{tikzpicture}
\caption{$k<0$ case. There are three types depending on $\nu_1,\nu_2$ and $r$. We denote the height to  $x_3$ direction by the number written in each rectangle.\label{fig:identification2}}
\end{center}
\end{figure}

We move to analyze the character. As in the $n=1$ case, the Hilbert space in $Y_{1,2,0}$ can be identified with the plane partition with an extra pit at $(1,1,n+1)$ which implies the translation rule discussed in subsection \ref{subsec:truncation}. If the $U(1)$ weight of intermediate Young diagram is given by $\mu=kn+r\ (k\in\mathbb{Z},0\leq r<n)$, the plane partition can be identified as shown in Figure \ref{fig:identification1}, \ref{fig:identification2}:
\begin{figure}[H]
\begin{center}
\begin{tikzpicture}
\footnotesize

\draw[->] (0,0)--(0,-3.7);
\draw[->] (0,0)--(5.3,0);

\draw (0,-0.5)--(5,-0.5);
\draw (1,-0.5)--(1,-1)--(5,-1);
\draw (3,-1.5)--(3,-2)--(5,-2);
\draw (4,-2)--(4,-2.5)--(5,-2.5);
\draw (0.5,-0.5)--(0.5,-3.5);
\draw (1,-1)--(1,-3.5);

\node (3) at (5.6,0) {$2$};
\node (2) at (0,-4) {$1$};
\node (dot) at (4,-1.2) {$\vdots$};

\fill[pattern=north east lines, opacity=0.5] (0,0)--(0,-0.5)--(5,-0.5)--(5,0)--cycle;
\fill[pattern=north east lines, opacity=0.5] (1,-0.5)--(1,-1)--(5,-1)--(5,-0.5)--cycle;
\fill[pattern=north east lines, opacity=0.5] (3,-1.5)--(3,-2)--(5,-2)--(5,-1.5)--cycle;
\fill[pattern=north east lines, opacity=0.5] (4,-2)--(4,-2.5)--(5,-2.5)--(5,-2)--cycle;
\fill[pattern=north east lines, opacity=0.5] (0,-0.5)--(0.5,-0.5)--(0.5,-3.5)--(0,-3.5)--cycle;
\fill[pattern=north east lines, opacity=0.5] (0.5,-0.5)--(1,-0.5)--(1,-3.5)--(0.5,-3.5)--cycle;

\draw (5,-2) arc(270:340:0.3 and 0.6);
\draw (5,0) arc(90:20:0.3 and 0.6);
\node (m) at (5.4,-1) {$k$};


\node (n1) at (2.5,-0.3) {$n$};
\node (n2) at (3.1,-0.8) {$n$};
\node (n3) at (4,-1.75) {$n$};
\node (n4) at (4.5,-2.3) {$r$};
\node (nu1) at (0.25,-2) {$\nu_1$};
\node (nu2) at (0.75,-2) {$\nu_2$};

\end{tikzpicture}
\caption{$k\geq0$ case. We denote the height to $x_3$ direction by the number written in each rectangle.\label{fig:identification1}}
\end{center}
\end{figure}
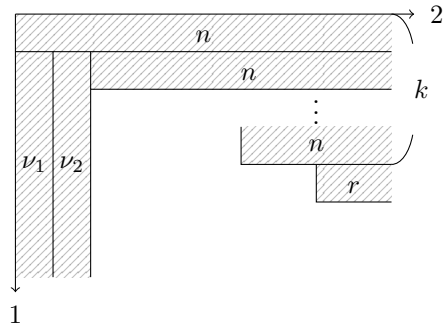

In a similar manner with the previous case, we need to add some boxes for the state to be not null. If $k>0$, the number of necessary boxes is 
\ba 
\sum_{i=1}^{k-1}\Bigl(2n(i-1)-(\nu_1+\nu_2)\Bigr)+(2k-2)r+(r-\nu_1)\theta(r>\nu_1)+(r-\nu_2)\theta(r>\nu_2),
\ea
where $\theta(P)$ is equal to 1 if $P$ is true and 0 if false. The last three terms correspond to boxes inserted in $(k+1)$th row. In the following, we set for simplicity
\ba
\theta=(r-\nu_1)\theta(r>\nu_1)+(r-\nu_2)\theta(r>\nu_2).
\ea
After adding boxes, the conformal dimension and $U(1)$ charge (including the contribution from $Y_{1,0,0}$ side and decoupling $U(1)$ factor) are computed from (\ref{eq:totaldim}) and the above discussion as
\ba
\begin{split}
&h=\frac{-2\nu_1\nu_2+\nu_1+(2n+3)\nu_2}{2(n+2)}+\frac{n(n+2)}{2}k^2+\Bigl((n+2)r-(\nu_1+\nu_2)\Bigr)k
+\frac{r^2}{2}-r+\theta,\\
&J=kn+r-\frac{\nu_1+\nu_2}{n+2}.
\end{split}
\ea
Note that we need to consider the overlapping boxes in the above computation.  One can check that this expression is also true of $k<0$ case. In terms of the parameters $l,m$ introduced in $(\ref{eq:n2primary})$, it can be  rewritten  as follows:
\ba
\label{eq:619}
\begin{cases}
&h=\frac{n(n+2)}{2}\Bigl(k+\frac{m+(r-1)(n+2)}{n(n+2)}\Bigr)^2+\frac{l(l+2)}{4(n+2)}-\frac{(m+2(r-1))^2}{4n}+\theta\qquad\raisebox{-18pt}{($\nu_2\geq1$, or  $l\neq-m$),}\\
&J=\frac{m}{n+2}+kn+r-1\\
\ \\
&h=\frac{n(n+2)}{2}\Bigl(k+\frac{m+r(n+2)}{n(n+2)}\Bigr)^2+\frac{l(l+2)}{4(n+2)}-\frac{(-l+2r)^2}{4n}+(r-l)\theta(r>l)\hspace{50pt}\raisebox{-18pt}{($\nu_2=0$, or  $l=-m$).}\\
&J=\frac{m}{n+2}+kn+r
\end{cases}
\ea
We can obtain the character by multiplying the factor from the highest weight (\ref{eq:619}) by the character of the plane partition obtained after the process of adding boxes. 
To do that, we need to know the explicit expression for the plane partition's character\footnote{We note that we need to consider the periodicity rule here.}. As we have already seen,  it corresponds to the character for some degenerate module of $W_n$ ($+ U(1)$) minimal model up to the factor of the highest weight.  In the language of section \ref{sec:w-minimal}, it is parametrized by $p=n+1$ and $q=n+2$. One can also see it as the character of 
para-fermion ($+ U(1)$ factor)\footnote{The relation between  $W$ algebra and para-fermion was studied in \cite{Quijano1,Quijano2}.}. Let's clarify which module  corresponds to  the given asymptotic Young diagram from the viewpoint of para-fermion. We first note that we can adjust the asymptotic Young diagram in $x_2$ direction to trivial one; one can change the order of Young diagrams in Figure \ref{fig:minimalW} cyclically so that  $\tilde{n}_i'=0$ for $i=1,\cdots n-1$, so we only have to consider the asymptotic condition in $x_1$ direction.
When the asymptotic Young diagram is parametrized by $(l,m)$ in the way of (\ref{eq:mapping}),
its conformal dimension (after decoupling $U(1)$ factor) is, 
\ba
\label{eq:paraweight}
h=\begin{cases}
&h_{l,m-2}^{\rm PF}\quad(l\neq-m)\\
&h_{l,m}^{\rm PF}\qquad(l=-m)
\end{cases}
\ea
where 
\ba
\label{eq:paraferdim}
h_{l,m}^{\rm PF}=\frac{l(l+2)}{4(n+2)}-\frac{m^2}{4n}.
\ea
The character for the product of this module and $U(1)$ factor is given by string function $c_{l,m}^{(n)}(\tau)$ (see Appendix \ref{app:stringfunc} for the definition). To avoid confusion,  we denote the parameters associated with SCA by $(l,m)_{\rm SCA}$ and the one associated with para-fermion by $(l,m)_{\rm PF}$ in the following.

The remaining thing we have to do is to clarify to which module the plane partition obtained after adding boxes to Figure {\ref{fig:identification1}},{\ref{fig:identification2}}  corresponds.
If $k<0$, the configuration is, up to shift of the origin, the plane partition with the asymptotic Young diagram in $x_1$ direction with two rows whose length are $n\theta(\nu_1<r)+\nu_1-r$ and $n\theta(\nu_2<r)+\nu_2-r$ (see Figure \ref{fig:identification2}). This expression is  also true of the case with $k>0$ after adjusting the asymptotic condition in $x_2$ direction. Changing the parameter from $(\nu_1,\nu_2)$ to $(l,m)_{\rm SCA}$,  the parameter of the corresponding module is read off from (\ref{eq:mapping}) and (\ref{eq:paraweight}). The result in the case of $l\neq-m\ (\nu_2\neq0)$ is as follows:
\ba
\label{eq:parafer}
\begin{cases}
(l,m+2(r-1))_{\rm PF}\hspace{65pt} (r\leq\nu_2)\\ 
(n-l,m+2(r-1)-n)_{\rm PF}\qquad (\nu_2<r\leq\nu_1)\\ 
(l,m+2(r-1)-2n)_{\rm PF}\hspace{40pt} (r>\nu_1).
\end{cases}
\ea
Following the convention where the region of the parameter $m$ is extended to $m\in\mathbb{Z}_{2n}$ by the identification 
\ba
\label{eq:paraidentification}
(n-l,m+n)_{\rm PF}\equiv(l,m)_{\rm PF},\quad(l,m+2n)_{\rm PF}\equiv(l,m)_{\rm PF},
\ea
the result (\ref{eq:parafer}) can be unified in the single form $(l,m+2(r-1))_{\rm PF}$.
After short computation, one can also see that the highest weight of these modules can be represented  in the single form as 
\ba
h_{l,m+2(r-1)}^{\rm PF}+\theta.
\ea
This factor appears in (\ref{eq:619}) and the product with the plane partition's character gives the string function $c_{l,m+2(r-1)}^{(n)}$. Given the above results, we have the character as follows:
\ba
\begin{split}
\chi(\tau,z)&=\sum_{r\in\mathbb{Z}_n}c_{l,m+2(r-1)}^{(n)}(\tau)\sum_{k=-\infty}^{\infty}q^{\frac{n(n+2)}{2}\bigl(k+\frac{m+(r-1)(n+2)}{n(n+2)}\bigr)^2}y^{kn+r-1+\frac{m}{n+2}}\\
&=\sum_{r\in\mathbb{Z}_n}c_{l,m+2r}^{(n)}(\tau)\bigl(\Theta_{2m+2r(n+2),2n(n+2)}(\tau,\frac{z}{n+2})+\Theta_{2m+2(r+n)(n+2),2n(n+2)}(\tau,\frac{z}{n+2})\bigr)\\
&=\sum_{r\in\mathbb{Z}_{2n}}c_{l,m+2r}^{(n)}(\tau)\Theta_{2m+2r(n+2),2n(n+2)}(\tau,\frac{z}{n+2}),
\end{split}
\ea
which is consistent with the literature. One can check it is also true of $l=-m$ case in a similar manner.

\section{Conclusion}
In this paper, we demonstrate that the double-truncation of the plane partition reproduces the minimal models of $W$-algebra and the simplest example of WoW, $\mathcal{N}=2$ superconformal algebra. One may conjecture that the double truncation is a universal method to describe the minimal models of the VOA family obtained as WoW. It will be interesting to explore the other WoW VOAs proposed in \cite{Prochazka:2017qum} whose minimal models are not well-known. To study the higher rank $\mathcal{N}=2$ super W-algebra \cite{Gaberdiel:2017hcn, Gaberdiel:2018nbs} may be other direction. We suppose that to clarify the consistency with the modular property will be essential in the further steps. 

As we have seen, the construction of WoW by PP remains somewhat mysterious.  The direction of the shared infinite leg and the shape of the intermediate Young diagram depend on the $U(1)$ charge of the intermediate channel while we may interpret it as an analytic continuation. Such complication will be more serious when we consider more involved diagrams. The operation of the extended generators such as proposed in \cite{Gaberdiel:2018nbs} is necessary while the way to include the negative charge diagrams seems different.

The construction of WoW reminds us of the computation of the topological string amplitude written by the topological vertex \cite{Aganagic:2003db}. The summation in the intermediate Young diagram looks similar to the composition of topological vertices. While there seems to be some difference in the treatment of negative charges, we hope to see some implications in the relation between WoW and, for example, the quiver W-algebra \cite{Kimura:2015rgi} and the relation with AFS vertex \cite{Awata:2011ce, Bourgine:2017jsi}.

\section*{Acknowledgement}
KH would like to thank M.Fukuda for helpful comments. 
YM appreciates discussion with Hong Zhang. He also thanks comments from the participants of the workshop and school ``Topological Field Theories, String theory and Matrix Models - 2018" (August 20 - August 25, 2018, Moscow) where the preliminary result of paper was presented. The research of YM is partially supported by Grant-in-Aid MEXT/JSPS KAKENHI 18K03610. KH is supported in part by JSPS fellowship.

\appendix

\section{The conformal dimension of the primary field in W-algebra minimal models from affine Yangian}
\label{app:Wdim}
The conformal dimension $h$ is computed  in the manner explained in section \ref{subsubsec:degeneratemodule}. It is done by four steps as follows. The first factor comes from the Young diagram $\mu$ and its contribution can be read off from (\ref{eq:formula}) as 
\ba
\begin{split}
h_1&=-\frac{\lambda_1}{2\lambda_2}\sum_{i=1}^{N-1}\mu_i^2-\frac{\lambda_1}{2\lambda_3}\sum_{i=1}^{N-1}(2i-1)\mu_i+\frac{\lambda_1}{2}\sum_{i=1}^{N-1}\mu_i\\
&=\frac{\beta}{2}\sum_{i=1}^{N-1}\mu_i^2+\frac{1-\beta}{2}\sum_{i=1}^{N-1}(2i-1-N)\mu_i.
\end{split}
\ea
In the same way, we have the second one coming from the Young diagram $\mu'$ as
\ba
h_2=\frac{1}{2\beta}\sum_{i=1}^{N-1}\mu_i'^2-\frac{1-\beta}{2\beta}\sum_{i=1}^{N-1}(2i-1-N)\mu_i'.
\ea
As we explained in section \ref{subsubsec:degeneratemodule}, summing the above two factors is not enough to get the conformal dimension. We also need to subtract the number of overlapping boxes 
\ba
\#(\mu\cap\mu')=\sum_{i=1}^{N-1}\mu_i\mu_i'
\ea
from it.
Finally, we have to decouple $U(1)$ factor, whose Virasoro zero mode corresponds to $\frac{\psi_1^2}{2\psi_0}$. Using (\ref{eq:formula}), we have 
\ba
\begin{split}
h^{U(1)}&=\frac{1}{2\psi_0}\Bigl(-\frac{\sum_{i=1}^{N-1}\mu_i}{h_1}-\frac{\sum_{i=1}^{N-1}\mu_i'}{h_2}\Bigr)^2\\
&=\frac{\beta}{2N}\bigl(\sum_{i=1}^{N-1}\mu_i\bigr)^2+\frac{1}{2N\beta}\bigl(\sum_{i=1}^{N-1}\mu_i'\bigr)^2-\frac{1}{N}\bigl(\sum_{i=1}^{N-1}\mu_i\bigr)\bigl(\sum_{i=1}^{N-1}\mu_i'\bigr).
\end{split}
\ea
Here, we use the relation between the parameters such as $\frac{1}{\psi_0h_1^2}=\frac{h_2}{\psi_0\sigma_3}\frac{h_3}{h_1}=-\frac{\lambda_1}{\lambda_2\lambda_3}$. 
Combining the above results, we have
\ba
\label{eq:Wweight}
\begin{split}
h&=h_1+h_2-\#(\mu\cap\mu')-h^{U(1)}\\
&=\frac{N-1}{2N}\bigl(\beta\sum_{i=1}^{N-1}\mu_i^2-2\sum_{i=1}^{N-1}\mu_i\mu_i'+\frac{1}{\beta}\sum_{i=1}^{N-1}\mu_i'^2\bigr)
-\frac{1}{2N}\bigl(\beta\sum_{i\neq j}\mu_i\mu_j-2\sum_{i\neq j}\mu_i\mu_j'+\frac{1}{\beta}\sum_{i\neq j}\mu_i'\mu_j'\bigr)\\
&\hspace{280pt}+\frac{1-\beta}{2}\sum_{i=1}^{N-1}(2i-1-N)(\mu_i-\frac{1}{\beta}\mu_j')\\
&=\frac{1}{2pq}\biggl(\frac{N-1}{N}\sum_{i=1}^{N-1}(p\mu_i-q\mu_i')^2-\frac{1}{N}\sum_{i\neq j}(p\mu_i-q\mu_i')(p\mu_j-q\mu_j')+(p-q)\sum_{i=1}^{N-1}(N+1-2i)(p\mu_i-q\mu_i')\biggr).
\end{split}
\ea
To compare it with (\ref{eq:cdim}), we introduce the weights for the defining representation of $su(N)$ and denote them by $\vec{\nu}_i\  (i=1\cdots N)$. They satisfy the following property:
\ba
\begin{split}
&\vec{\nu}_i\cdot\vec{\nu}_j=-\frac{1}{N}+\delta_{i,j},\quad\vec{\nu}_i\cdot\vec{\rho}=\frac{N+1}{2}-i,\\
&\omega_i=\sum_{j=1}^i\nu_j\quad (i=1\cdots N-1), 
\end{split}
\ea
where $\vec{\rho}=\sum_{I=1}^{N-1}\vec{\omega}_i$ is a Weyl vector. Using them, we can rewrite (\ref{eq:Wweight}) as 
\ba
\begin{split}
h&=\frac{1}{2pq}\biggl(\Bigl(\sum_{i=1}^{N-1}(p\mu_i-q\mu_i')\vec{\nu}_i\Bigr)^2+2(p-q)\sum_{i=1}^{N-1}(p\mu_i-q\mu_i')\vec{\nu}_i\cdot\vec{\rho}\biggr)\\
&=\frac{1}{2pq}\biggl(\Bigl(\sum_{i=1}^{N-1}(p\mu_i-q\mu_i')\vec{\nu}_i+(p-q)\vec{\rho}\Bigr)^2-(p-q)^2\frac{N(N^2-1)}{12}\biggr)\\
&=\frac{12(\sum_{i=1}^{N-1}(pn_i-qn'_i)\vec \omega_i)^2-N(N^2-1)(p-q)^2}{24pq},
\end{split}
\ea
which is consistent with (\ref{eq:cdim}). Here, we use the formula $\vec{\rho}\cdot\vec{\rho}=\frac{N(N^2-1)}{12}$.

\section{The norm of the primary state composed of supercurrent}
\label{app:a}
$\mathcal{N}=2$ super Virasoro algebra is defined as follows:
\begin{equation}
\begin{split}
&[L_n,L_m]=(m-n)L_{n+m}+\frac{c}{12}(m^3-m)\delta_{n+m,0},\\
&[L_m,G^\pm_r]=(\frac{1}{2}m-r)G_{m+r}^\pm,\\
&[L_n,J_n]=-nJ_{n+m},\\
&[J_m,J_n]=\frac{c}{3}m\delta_{n+m,0},\\
&[J_m,G_r^\pm]=\pm G_{m+r}^\pm,\\
&\{G_r^+,G_s^-\}=2L_{r+s}+(r-s)J_{r+s}+\frac{c}{3}(r^2-\frac{1}{4})\delta_{r+s,0},\\
&\{G_r^\pm,G_s^\pm\}=0.
\end{split}
\end{equation}
Using it, we can compute the norm of the intermediate Young diagram with $U(1)$ weight $\mu$ which corresponds to the primary state (\ref{eq:supercurrentprimary}):
\ba
\begin{split}
&\bra{0}G^{\mp}_{3/2}\cdots G^{\mp}_{|\mu|-1/2}G^{\mp}_{|\mu|+1/2}G^{\pm}_{-|\mu|-1/2}G^{\pm}_{-|\mu|+1/2}\cdots G^{\pm}_{-3/2}\ket{0}\\
=&\bra{0}G^{\mp}_{3/2}\cdots G^{\mp}_{|\mu|-1/2}\{G^{\mp}_{|\mu|+1/2}, G^{\pm}_{-|\mu|-1/2}\}G^{\pm}_{-|\mu|+1/2}\cdots G^{\pm}_{-3/2}\ket{0}\\
=&\bra{0}G^{\mp}_{3/2}\cdots G^{\mp}_{|\mu|-1/2}\Bigl(2L_0\mp2\bigl(|\mu|+\frac{1}{2}\bigr)J_0+\frac{c}{3}\bigl((|\mu|+\frac{1}{2})^2-\frac{1}{4}\bigr)\Bigr)G^{\pm}_{-|\mu|+1/2}\cdots G^{\pm}_{-3/2}\ket{0}\\
=&\frac{|\mu|^2+|\mu|}{3}\Bigl(c-\frac{3(|\mu|-1)}{|\mu|+1}\Bigr)\bra{0}G^{\mp}_{3/2}\cdots G^{\mp}_{|\mu|-1/2}G^{\pm}_{-|\mu|+1/2}\cdots G^{\pm}_{-3/2}\ket{0}\\
=&\cdots\\
=&\prod_{i=1}^{|\mu|}\frac{i(i+1)}{3}\Bigl(c-\frac{3(i-1)}{i+1}\Bigr)
\end{split}
\ea
Note that the other terms with anti-commutator which should appear in the second line vanish.  The result shows that if $c=\frac{3n}{n+2}$ ($n\in\mathbb{N}$), the above states with $|\mu|>n$ become null.

\section{The character of parafermion}
\label{app:stringfunc}
$\mathbb{Z}_n$ parafermion is defined by the coset $\frac{\widehat{SU}(2)_n}{\widehat{U}(1)}$. Here, we denote $SU(2)$ affine Kac-Moody algebra with level $n$ by $\widehat{SU}(2)_n$. We use the following normalization:
\ba
J^a(z)J^b(w)\sim\frac{\frac{n}{2}\delta_{ab}}{(z-w)^2}+\frac{i\epsilon^{abc}J^c(w)}{z-w}.
\ea
The integrable representation of $\widehat{SU}(2)_n$  is parametrized by the eigenvalue  of $J_0^3$ as 
\ba
J_m^a\ket{l}=0\ (m>0),\quad J_0^+\ket{l}=0,\quad J_0^3\ket{l}=\frac{l}{2}\ket{l}.
\ea
The parameter $l$  can take the integer value satisfying $0\leq l\leq n$. In each value of $l$, there are several primary fields $\ket{l,m}\propto(J_0^-)^{\frac{l-m}{2}}\ket{l}$ for $-l\leq m\leq l$ and $l-m\equiv 0 \ ({\rm mod} 2)$. It is decomposed into the primary field of parafermion and that of $\widehat{U}(1)$. The conformal dimension of the primary field for parafermion parametrized by $(l,m)$ can be computed by Sugawara construction and then obtain (\ref{eq:paraferdim}). There are highest weight states for $\widehat{U}(1)$ also in the descendant of $\widehat{SU}(2)_n$. If  the state has $\widehat{U}(1)$ charge $\frac{j}{2}$ satisfying $j\equiv m$ (mod $n$), the parafermion parametrized by $(l,m)$ acts on it. This should be interpreted in the meaning of (\ref{eq:paraidentification}) if $m$ takes the value out of the region $-l\leq m\leq l$. The character can be computed by decomposing $\widehat{SU}(2)_n$ character $\chi^{SU(2)}_{l,n}(\tau,z):={\rm Tr}\ q^{L_0-\frac{c}{24}}y^{J_0^3}$ ($q={\rm e}^{2\pi i\tau},y={\rm e}^{2\pi iz}$) into the parafermion character $\chi_{l,m}^{\rm PF}(\tau)$ and $U(1)$ character as follows:
\ba
\chi_{l,n}^{SU(2)}(\tau,z)=\sum_{m\in\mathbb{Z}_n}\chi_{l,m}^{\rm PF}(\tau)\frac{\Theta_{m,n}(\tau,z)}{\eta(\tau)},
\ea
where
\ba
\label{eq:etatheta}
\eta(\tau)=q^{\frac{1}{24}}\prod_{i=1}^{\infty}(1-q^i),\quad\Theta_{m,n}(\tau,z)=\sum_{k=-\infty}^{\infty}q^{n(k+\frac{m}{2n})^2}y^{n(k+\frac{m}{2n})}.
\ea
String function is defined by 
\ba
\chi_{l,n}^{SU(2)}(\tau,z)=\sum_{m\in\mathbb{Z}_n}c_{l,m}^{(n)}(\tau)\Theta_{m,n}(\tau,z).
\ea
It corresponds to the character for the product of parafermion and $U(1)$ factor. As is expected from (\ref{eq:paraidentification}), it satisfies 
\ba
c_{l,m+2n}^{(n)}(\tau)=c_{n-l,m+n}^{(n)}(\tau)=c_{l,m}^{(n)}(\tau).
\ea
Ww note that string function is  equal to $\frac{1}{\eta(\tau)}$ when $n=1$.

\bibliography{WoW}

\end{document}